\documentclass[twocolumn]{aastex62}
\usepackage{dcolumn}
\usepackage{multirow}
\usepackage{rotating}
\usepackage{tabularx}
\received{July 30, 2019}
\revised{xxx} 
\accepted{xxx}
\submitjournal{ApJ}

\shorttitle{Internal dynamics and stellar content of 9 UDGs}
\shortauthors{Chilingarian et al.}

\begin{document}

\title{Internal dynamics and stellar content of nine ultra-diffuse galaxies in the Coma cluster prove their evolutionary link with dwarf early-type galaxies\footnote{All galaxy spectra presented in this work and their best-fitting stellar population templates are available at the following permanent address: \url{https://doi.org/10.5281/zenodo.3387745}}}


\correspondingauthor{Igor V. Chilingarian}
\email{igor.chilingarian@cfa.harvard.edu, chil@sai.msu.ru}

\author[0000-0002-7924-3253]{Igor V. Chilingarian}
\affiliation{Smithsonian Astrophysical Observatory, 60 Garden St. MS09, Cambridge, MA 02138, USA}
\affiliation{Sternberg Astronomical Institute, M.V.Lomonosov Moscow State University, Universitetsky prospect 13, Moscow, 119234, Russia}

\author[0000-0002-8220-0756]{Anton V. Afanasiev}
\affiliation{Sternberg Astronomical Institute, M.V.Lomonosov Moscow State University, Universitetsky prospect 13, Moscow, 119234, Russia}
\affiliation{Department of Physics, M.V. Lomonosov Moscow State University, 1 Vorobyovy Gory, Moscow, 119991, Russia}

\author[0000-0003-3255-7340]{Kirill A. Grishin}
\affiliation{Sternberg Astronomical Institute, M.V.Lomonosov Moscow State University, Universitetsky prospect 13, Moscow, 119234, Russia}
\affiliation{Department of Physics, M.V. Lomonosov Moscow State University, 1 Vorobyovy Gory, Moscow, 119991, Russia}

\author[0000-0002-1311-4942]{Daniel Fabricant}
\affiliation{Smithsonian Astrophysical Observatory, 60 Garden St. MS09, Cambridge, MA 02138, USA}

\author[0000-0002-9194-5071]{Sean Moran}
\affiliation{Smithsonian Astrophysical Observatory, 60 Garden St. MS09, Cambridge, MA 02138, USA}

\begin{abstract}
Ultra-diffuse galaxies (UDGs) are spatially extended, low surface brightness stellar systems with regular elliptical-like morphology found in a wide range of environments. Studies of the internal dynamics and dark matter content of UDGs that would elucidate their formation and evolution have been hampered by their low surface brightnesses. Here we present spatially resolved velocity profiles, stellar velocity dispersions, ages and metallicities for 9 UDGs in the Coma cluster.  We use intermediate-resolution spectra obtained with Binospec, the MMT's new high-throughput optical spectrograph. We derive dark matter fractions between 50~\%\ and 90~\% within the half-light radius using Jeans dynamical models. Three galaxies exhibit major axis rotation, two others have highly anisotropic stellar orbits, and one shows signs of triaxiality. In the Faber--Jackson and mass--metallicity relations, the 9 UDGs fill the gap between cluster dwarf elliptical (dE) and fainter dwarf spheroidal (dSph) galaxies. Overall, the observed properties of all 9 UDGs can be explained by a combination of internal processes (supernovae feedback) and environmental effects (ram-pressure stripping, interaction with neighbors). These observations suggest that UDGs and dEs are members of the same galaxy population.
\end{abstract}


\section{Introduction} \label{sec:intro}
Low-luminosity galaxies with regular morphologies and no current star formation represent the numerically dominant population in dense regions of the Universe \citep{SB84,FS88}.  These galaxies are commonly called dwarf ellipticals (dE) or dwarf lenticulars (dS0).  In contrast to giant early-type galaxies \citep{Kormendy77}, their central ($\mu_0$) and mean surface brightness within the half-light radius ($\langle \mu_e \rangle$) decrease at low luminosity. These dwarf galaxies are difficult targets for spectroscopy because $\langle \mu_e \rangle$ falls below the night sky brightness at luminosities $L_B < 10^9 L_{\odot}$ ($M_B>-17$~mag). These systems are usually not well resolved in cosmological numerical simulations due to their relatively small stellar masses ($10^7<M_*<5\cdot10^9 M_{\odot}$). The formation and evolution of dE/dS0s remain uncertain, although environmental effects in high density regions \citep[see  e.g.][]{1972ApJ...176....1G,1996Natur.379..613M} are thought to be responsible for the quenching of star formation and transformation of gas-rich disky progenitors into quiescent dE/dS0.

Recently, ``ultra-diffuse galaxies'' (UDGs) with larger sizes ($r_e = 1.5 - 4.5$~kpc) and lower central surface brightnesses ($\mu_{0g}>24.0$~mag/arcsec$^2$) than dE/dS0s were discovered \citep[][]{2015ApJ...798L..45V} in the Coma cluster. Soon after, \citet{Koda15} and \citet{yagi16} identified a large population of Coma UDGs in deep 8-m Subaru images adopting a different $r_e$ cut ($r_e>0.7$~kpc). These discoveries stimulated considerable interest in the origin and evolution of low-luminosity early-type galaxies. Some authors argued that UDGs are an extension of the dE/dS0 class \citep[see e.g.][]{2018RNAAS...2a..43C} with similar evolutionary histories. Others proposed that UDGs are formed by either early quenching of ``underdeveloped'' galaxies at $z\sim2$ \citep{2015MNRAS.452..937Y} or by strong star-formation driven outflows in dwarf galaxies \citep{2017MNRAS.466L...1D}.

Our understanding of UDGs is currently limited by the difficulty of obtaining high-quality spectra. Only about two dozen UDGs have even been confirmed as members of the clusters and groups where they were found photometrically \citep{2015ApJ...804L..26V,2017ApJ...838L..21K,FerreMateu+18,2018ApJ...859...37G,2018MNRAS.478.2034R}. Out of this sample, only about 15 allow an integrated light estimation of the UDG stellar population, and 6 provide [Fe/H] uncertainties better than 0.2~dex.  Just two UDGs have measured stellar velocity dispersions \citep{2016ApJ...828L...6V,Danieli+19}. Therefore, the internal dynamics and dark matter content of UDGs are poorly constrained.

Here we present the first results from a spectroscopic campaign that targeted 4 fields in the Coma cluster and 1 field in Abell~2147.  We present internal velocity dispersions, spatially resolved radial velocity profiles, and stellar population properties for 9 UDGs in the Coma cluster. Throughout this work we adopt a Coma cluster distance of 99~Mpc \citep{Saulder+16} with a corresponding distance modulus of $34.98$~mag and spatial scale of 0.5~kpc/arcsec.

\section{New Observations and Data Analysis} \label{sec:obs}

\subsection{Observations and Data Reduction}
\begin{figure*}
\includegraphics[width=0.97\hsize]{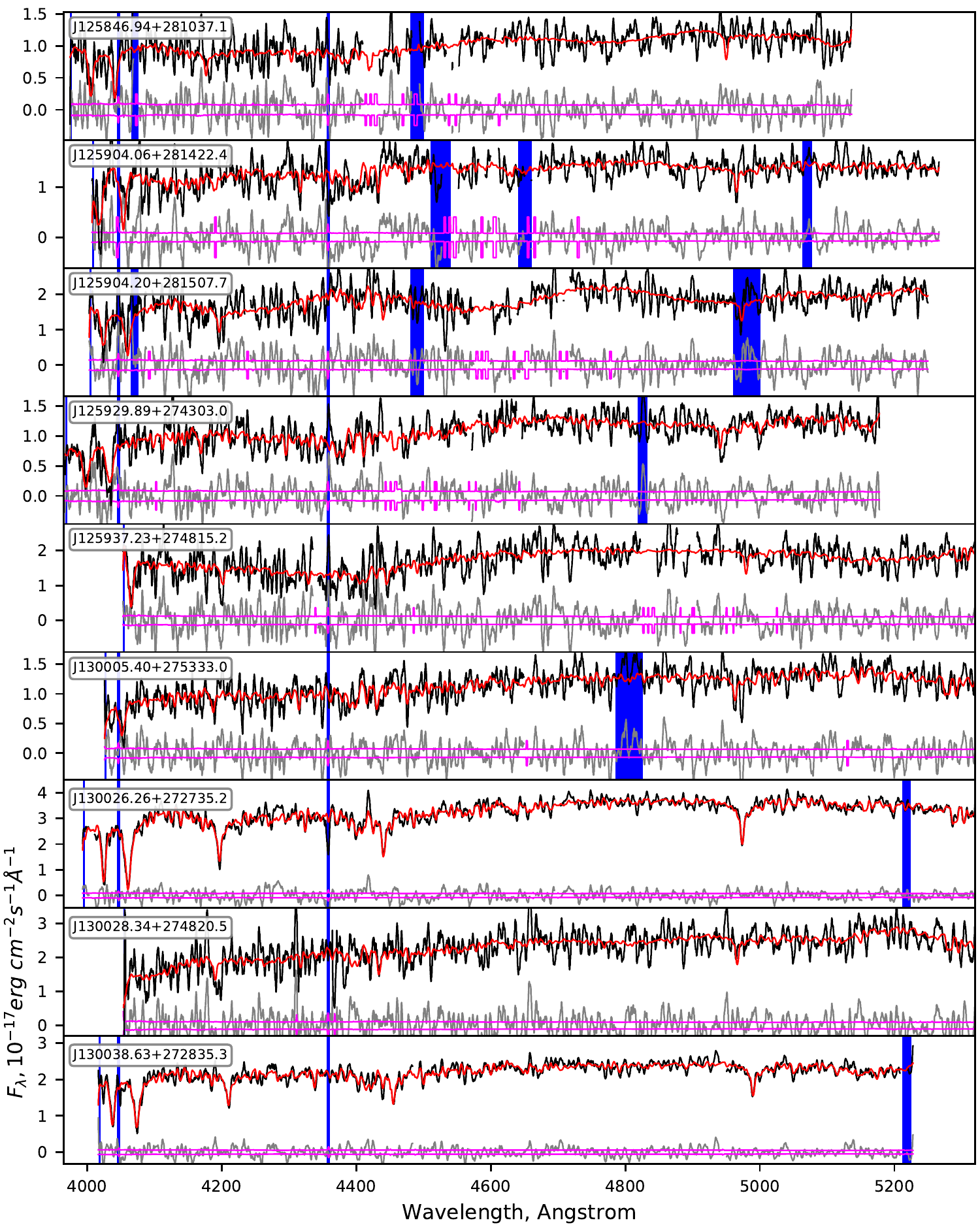}
\caption{Binospec spectra of nine UDGs optimally extracted along the slit flux calibrated in $F_{\lambda}$ units (black), their best-fitting stellar population models (red), flux uncertainties (purple), and fitting residuals (gray). Galaxy \emph{IAU}-style designations are shown in each panel. All data and best-fitting models are smoothed with the Savitzky-Golay filter for display purposes.\label{fig_udgspec}}
\end{figure*}

The UDG spectra we describe were obtained with the Binospec spectrograph commissioned in November 2017 at the 6.5m MMT \citep{2019PASP..131g5004F} We observed  three slit masks, starting in December 2017 during a shared-risk science verification program and continuing in April and June 2018 as normal queue programs. We selected targets for the first mask (``Coma1'') from the \citet{Adami08} photometric catalog. For the second (``ComaA'') and third (``ComaB'') masks we chose UDGs from the \citet{yagi16} sample limiting the effective surface brightness to $25.3> \langle \mu_{e,g} \rangle >24.4$~mag~arcsec$^{-2}$. We identified two galaxies, \emph{J130026.26+272735.2} and \emph{J130038.63+272835.3} in ``Coma1'' with similar properties to UDGs but with 3--4 times higher $\langle \mu_{e,R} \rangle$, slightly disturbed morphologies, and significantly bluer colors.  Both galaxies have axial ratios around 0.4--0.45 so that their deprojected surface brightness is $\approx$1~mag dimmer. Coupled with the low stellar $M/L$ ratios indicated by their blue colors, these galaxies certainly border the UDG regime, so we include them in our sample.

We used Binospec's 1000~gpm gratings with a $4500$~\AA\ central wavelength and with 1$^{\prime\prime}$-wide slitlets yielding wavelength coverage of 3,760 to 5,300~\AA\ and spectral resolution $R=3,750 \dots 4,900$ ($\sigma_{\mathrm{inst}} = 34 \dots 26$~km/s). The integration time was 2h~20min for ``Coma1'' and 2h for ``ComaA'' and ``ComaB'' split into 20min-long exposures. All three masks were observed under average seeing conditions (0.9--1.3$^{\prime\prime}$), good transparency in dark time on 2017/Dec/14 and 2018/Apr/10 (``Coma1''), 2018/Jun/4--5 (``ComaA''), and 2018/Jun/8 (``ComaB''). We took arc lamp and internal flat field frames at night time and high signal-to-noise sky flats in twilight to characterize the spectral resolution across the field and wavelength range.

We reduced the data with the Binospec pipeline\footnote{\url{https://bitbucket.org/chil_sai/binospec/src}} \citep{2019PASP..131g5005K} producing flux calibrated, sky-subtracted, rectified, and wavelength calibrated 2D long-slit images (with errors) for each slitlet. The image scale is 0.38~\AA\ per pixel in wavelength and 0.24$^{\prime\prime}$ per pixel along the slit. The slits are not aligned with the the major axes of the galaxies. We used non-local sky subtraction (i.e. a global sky model computed from all ``empty'' regions in all slits in the mask). The ``Coma1'' mask contained 10$^{\prime\prime}$-long slits centered on galaxies. In the ``ComaA'' and ``ComaB'' masks all the slits were extended where possible to improve modeling of the night sky background.

\subsection{Deriving Kinematics and Stellar Populations}
We analyzed the UDG spectra using the {\sc NBursts} technique \citep{CPSA07,CPSK07}, which implements a full spectrum fitting approach in pixel space using a grid of intermediate-spectral resolution ($R=10,000$) simple stellar population (SSP) models computed with the {\sc pegase.hr} evolutionary synthesis package \citep{LeBorgne+04}. We convolved SSPs with the Binospec spectral line-spread function determined from the analysis of twilight sky flats. As \citet{CCB08} demonstrated, the {\sc NBursts} algorithm can recover stellar velocity dispersions ($\sigma$) down to $\sigma_{\mathrm{inst}}/2$ at a signal-to-noise ratio of 5 per pixel. We convolved the stellar population templates with the line-of-sight velocity distribution (LOSVD) in Fourier space as described in \citet{cappellari17}, which substantially improved the stability of velocity dispersion measurements down to $0.3 \sigma_{\mathrm{inst}} \approx 9$~km/s.  Here we treat measurements below $\sigma=10$~km/s as upper limits. We use the restframe wavelength range between 3900\AA\ and 5200--5400\AA\ for the fitting procedure (the exact range depends on the slit position in the mask). See \citet{Chilingarian+08,Chilingarian09} for a detailed discussion regarding sensitivity, systematics, and limitations of the {\sc NBursts} full spectrum fitting. We allowed a 15th order multiplicative polynomial continuum correction and a constant additive term to account for uncertainties in subtracting scattered light. 

Velocity dispersion measurements from full spectrum fitting are subject to both random and systematic uncertainties. Random errors originate from flux uncertainties and depend on two principal factors, the signal-to-noise ratio per pixel and the average depth of absorption lines. The absorption line depth is affected by (a) spectral resolution and intrinsic velocity dispersion (lower resolution smears absorption features), (b) age and metallicity of the stellar population (metal-poor and/or young populations have weaker absorption lines), (c) the spectral wavelength range used in the analysis (different spectral regions of the same stellar population can be feature-rich, e.g. H$\beta$ and further blue or feature-poor, e.g. H$\alpha$ and further red). We address the velocity dispersion random errors in a separate paper (Chilingarian in prep.) where we conclude that poorer signal-to-noise ratio per pixel can be offset by  higher contrast absorption lines in a larger number of pixels. The most efficient measurements are made by choosing the optimal wavelength range and observing galaxies with low velocity dispersions at high spectral resolution. For example, a S/N=4 spectrum of a moderately metal-poor ([Fe/H]$=-1$~dex) old stellar population with an intrinsic $\sigma=20$~km/s observed in the 3900\AA--5300\AA~wavelength range at $R=4,500$ (Binospec) will yield $\Delta \sigma = 7$~km/s similar to a S/N=14 spectrum with an intrinsic $\sigma=44$~km/s observed around H$\alpha$ reported by \citet{2016ApJ...828L...6V}. In all studied cases, statistical errors of $\sigma$ reported by the non-linear minimization routine used inside {\sc nbursts} are fully consistent with the results of Monte-Carlo simulations even at low signal-to-noise levels (3--5 per pixel as in the spectra presented here).

The only source of systematic errors of velocity dispersion when fitting a purely Gaussian LOSVD to a spectrum with a perfectly known line-spread function is the template mismatch between the observed spectrum and the model. The most serious biases come from the degeneracy of velocity dispersion with the template spectrum metallicity or from poorly subtracted sky background \citep[see appendices in][for a detailed description]{CPSA07}. The high quality of the sky subtraction in our spectra is demonstrated by Poisson level residuals of bright atmospheric mercury lines (4047\AA\ and 4358\AA). Nevertheless, we included a zero-order additive term (i.e. a constant level) in the fitting to mitigate potential scattered light subtraction issues. To assess the systematic uncertainty arising from the $\sigma$--[Fe/H] degeneracy, we computed $\chi^2$ confidence maps by scanning the age--metallicity parameter space around the best-fitting solution and leaving $v, \sigma$ and polynomial multiplicative continuum terms as free parameters. The variance of $\sigma$ estimates at the 67\%-confidence level should provide an estimate of the velocity dispersion systematic error. At the low signal-to-noise ratio of our spectra, the $\sigma$ systematic errors are a factor of 2 or more below the random uncertainties.

In six cases we fit each slitlet spectrum twice.  We use an integrated spectrum using optimal extraction along the slit (see Fig.~\ref{fig_udgspec}) to best estimate the systemic velocity ($v_{\mathrm{sys}}$), stellar age ($t$), metallicity ([Fe/H]) and $\sigma$.  We assess radial velocity gradients using spectra binned along the slit to $S/N=1.5-2$ per pixel with stellar population parameters and $\sigma$ fixed at the values determined from the integrated spectrum. For \emph{J130005.40+275333.0} we fit $\sigma$ in 3 bins along the slit to minimize spectral line broadening due to the significant radial velocity gradient. We fit the $\sigma$ profile along the slit for the brighter galaxies \emph{J130026.26+272735.2} and \emph{J130038.63+272835.3}.  We present stellar kinematics of the 9 UDGs in Fig.~\ref{udg_models} in the middle panels.

\section{Archival Imaging Data and Jeans Anisotropic Modelling} \label{sec:dynmod}
\begin{figure*}
\centering
\begin{tabular}{rr}
\includegraphics[angle=0,width=0.35\hsize]{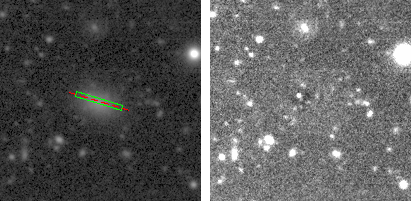} & 
\includegraphics[angle=0,width=0.35\hsize]{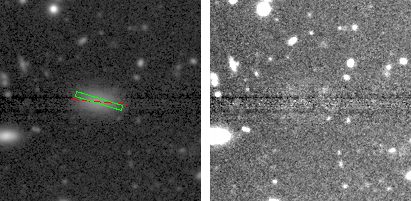} \\
\includegraphics[clip,trim={0cm 0 0.10cm 0.0cm}, width=0.41\hsize]{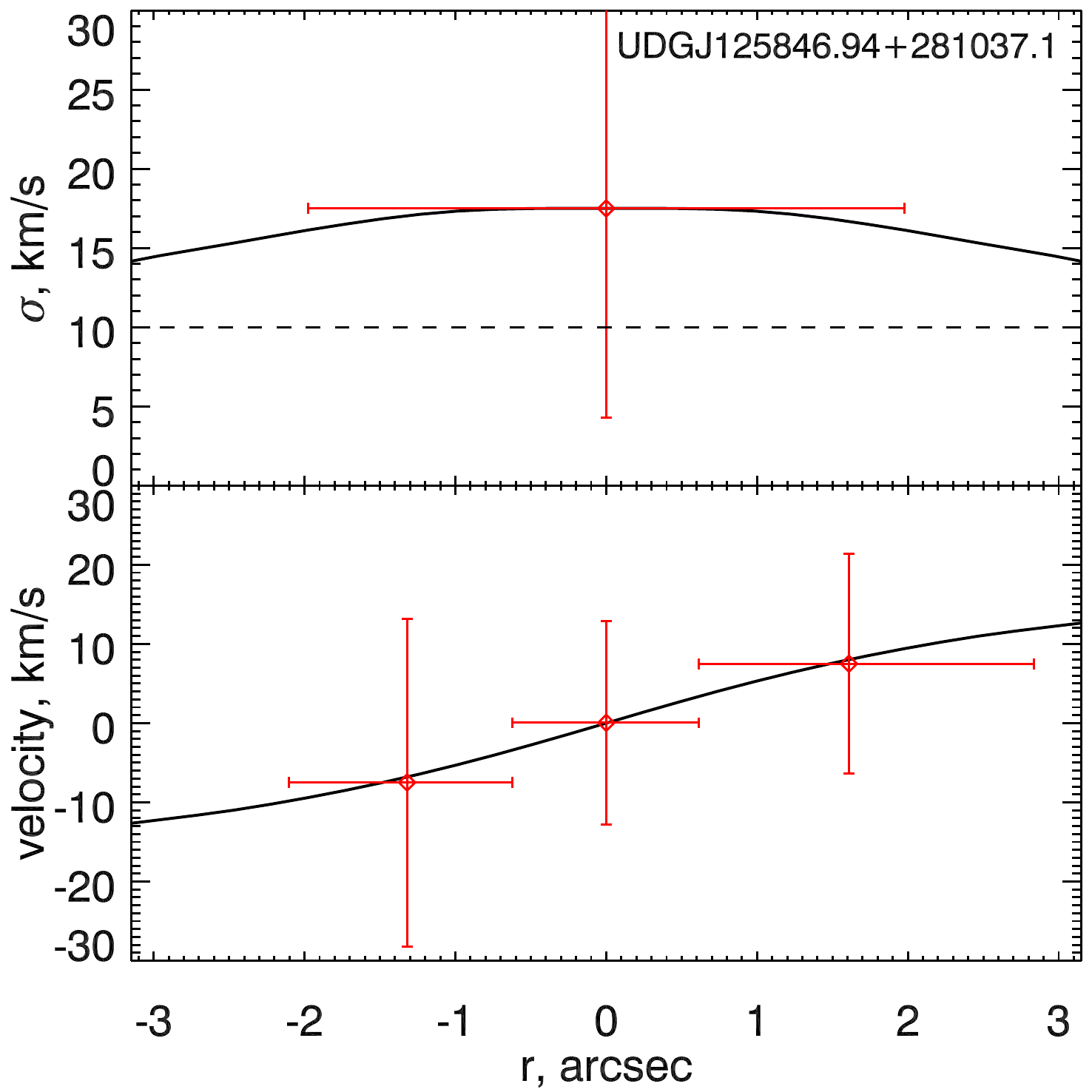} & 
\includegraphics[clip,trim={0cm 0 0.10cm 0.0cm}, width=0.41\hsize]{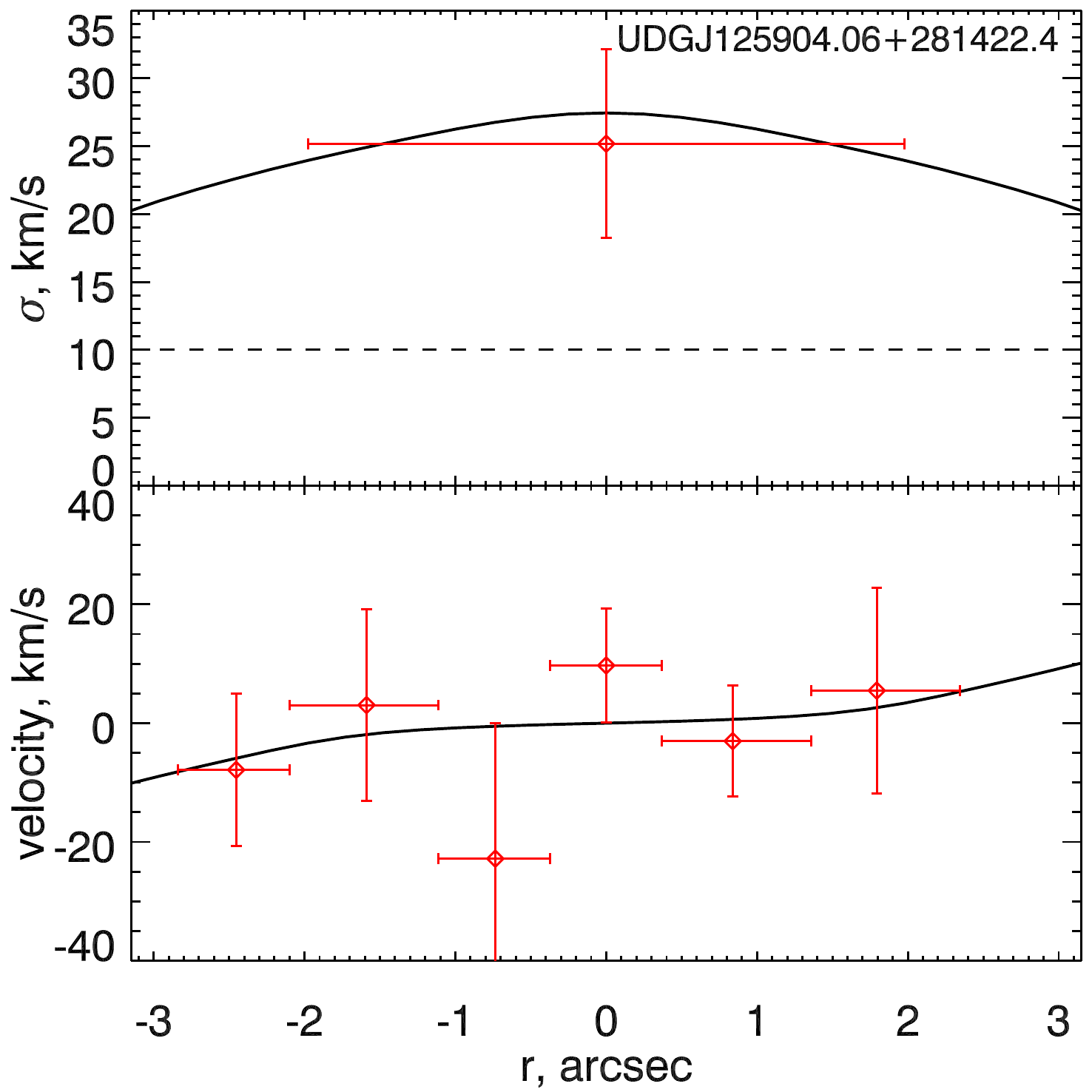} \\
\includegraphics[clip,trim={0.10cm 0 0.05cm 0.05cm}, width=0.41\hsize]{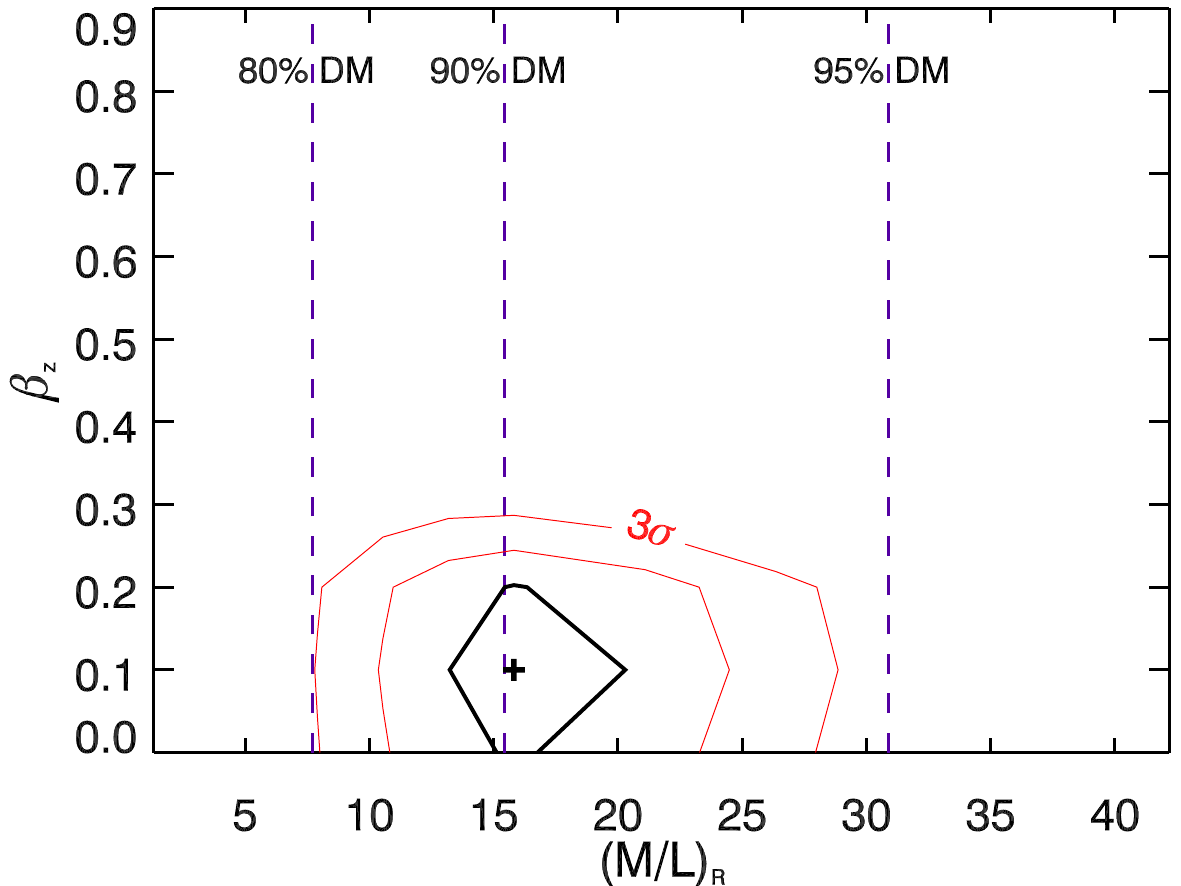} & 
\includegraphics[clip,trim={0.10cm 0 0.05cm 0.05cm}, width=0.41\hsize]{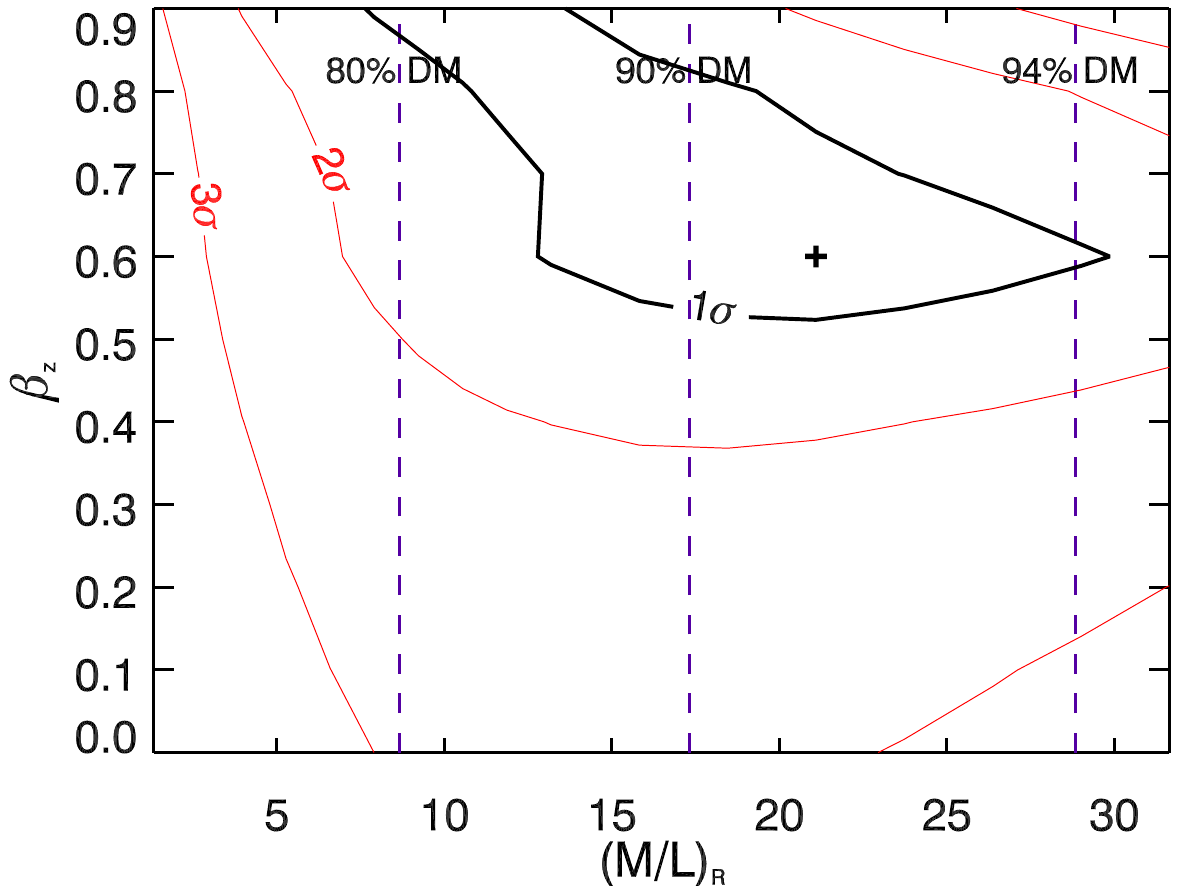} \\
\includegraphics[clip,trim={0.40cm 0.4cm 0.50cm 1.1cm}, width=0.425\hsize]{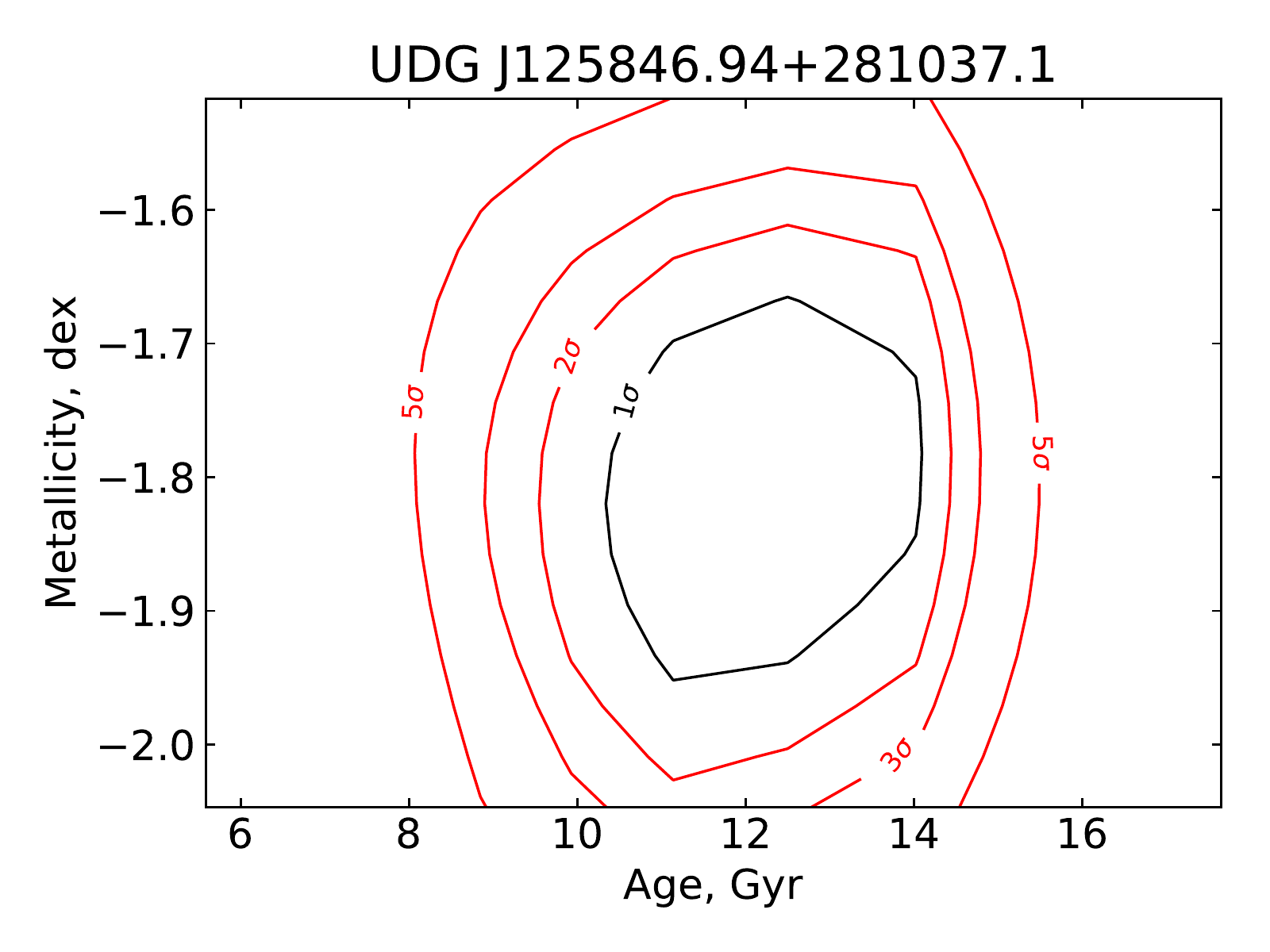} & 
\includegraphics[clip,trim={0.40cm 0.4cm 0.50cm 1.1cm}, width=0.425\hsize]{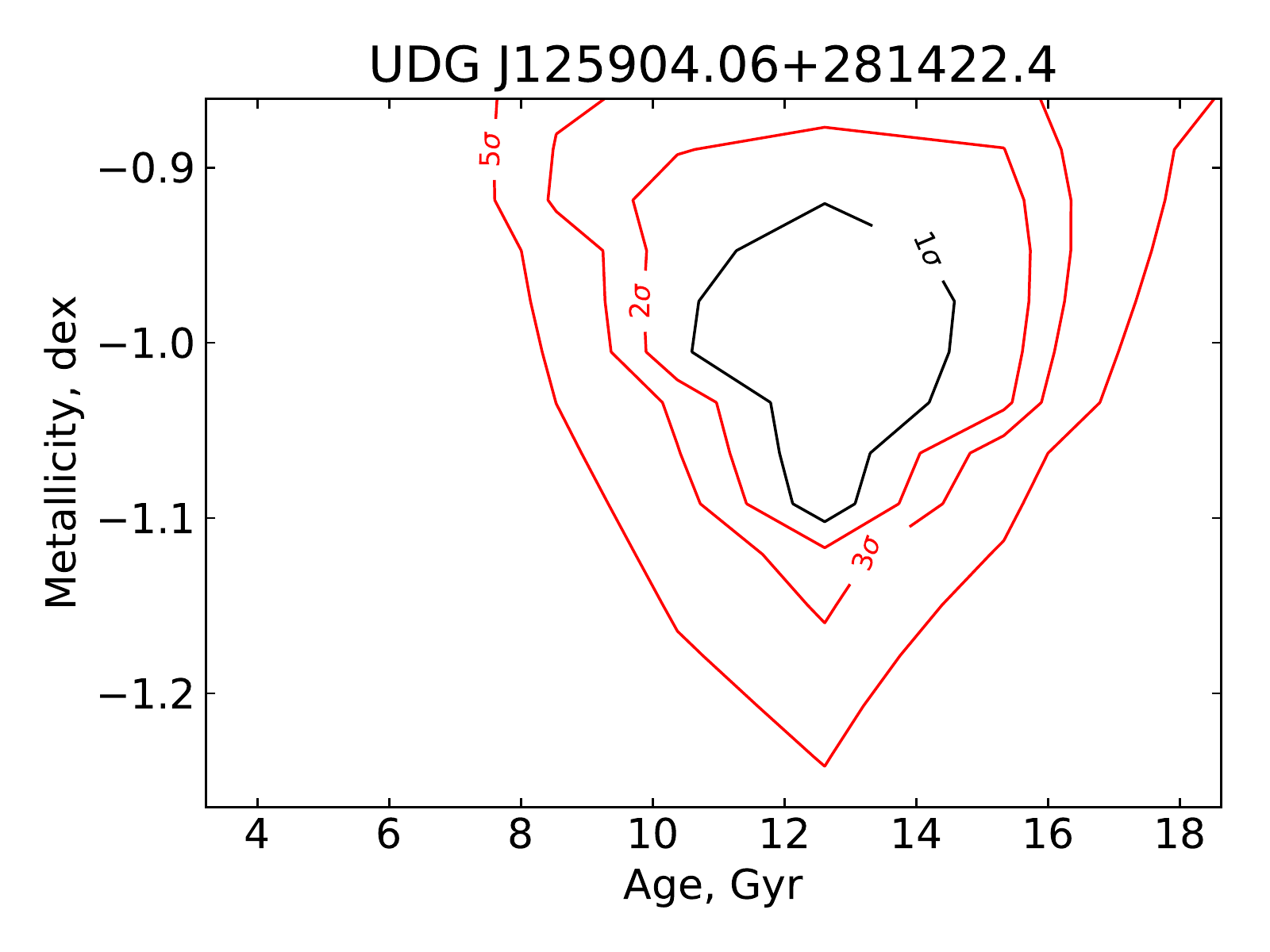}
\end{tabular}
\caption{Top row of panels: Subaru $R$-band 60$^{\prime\prime}$ $\times$60$^{\prime\prime}$ images (North is up; East is left) and residuals with their best-fitting {\sc galfit} models. Second row: stellar kinematics (top: velocity dispersion $\sigma$; bottom: radial velocity $v$). Third row: confidence maps in the $\beta_Z$ -- $M/L$ parameter space. Bottom row: confidence maps of stellar population parameters in the age--metallicity space. The 10$^{\prime\prime}$ $\times$1 $^{\prime\prime}$ Binospec slits are shown by green rectangles; a dashed red line indicates the major axis of each galaxy; black solid lines in the $v$ and $\sigma$ panels show the best-fitting Jeans models; and black dotted lines at $\sigma=$10~km/s show the adopted Binospec sensitivity limit.
\label{udg_models}}
\end{figure*}

\setcounter{figure}{1}
\begin{figure*}
\centering
\begin{tabular}{rr}
\includegraphics[angle=0,width=0.35\hsize]{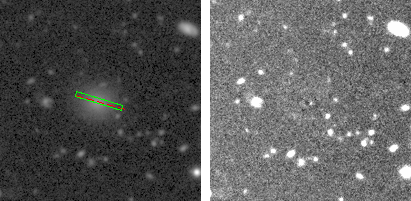} & 
\includegraphics[angle=0,width=0.35\hsize]{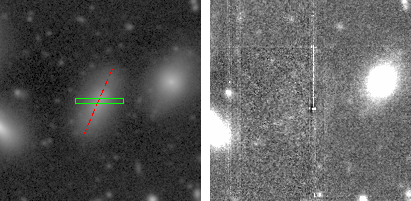} \\
\includegraphics[clip,trim={0cm 0 0.10cm 0.0cm}, width=0.41\hsize]{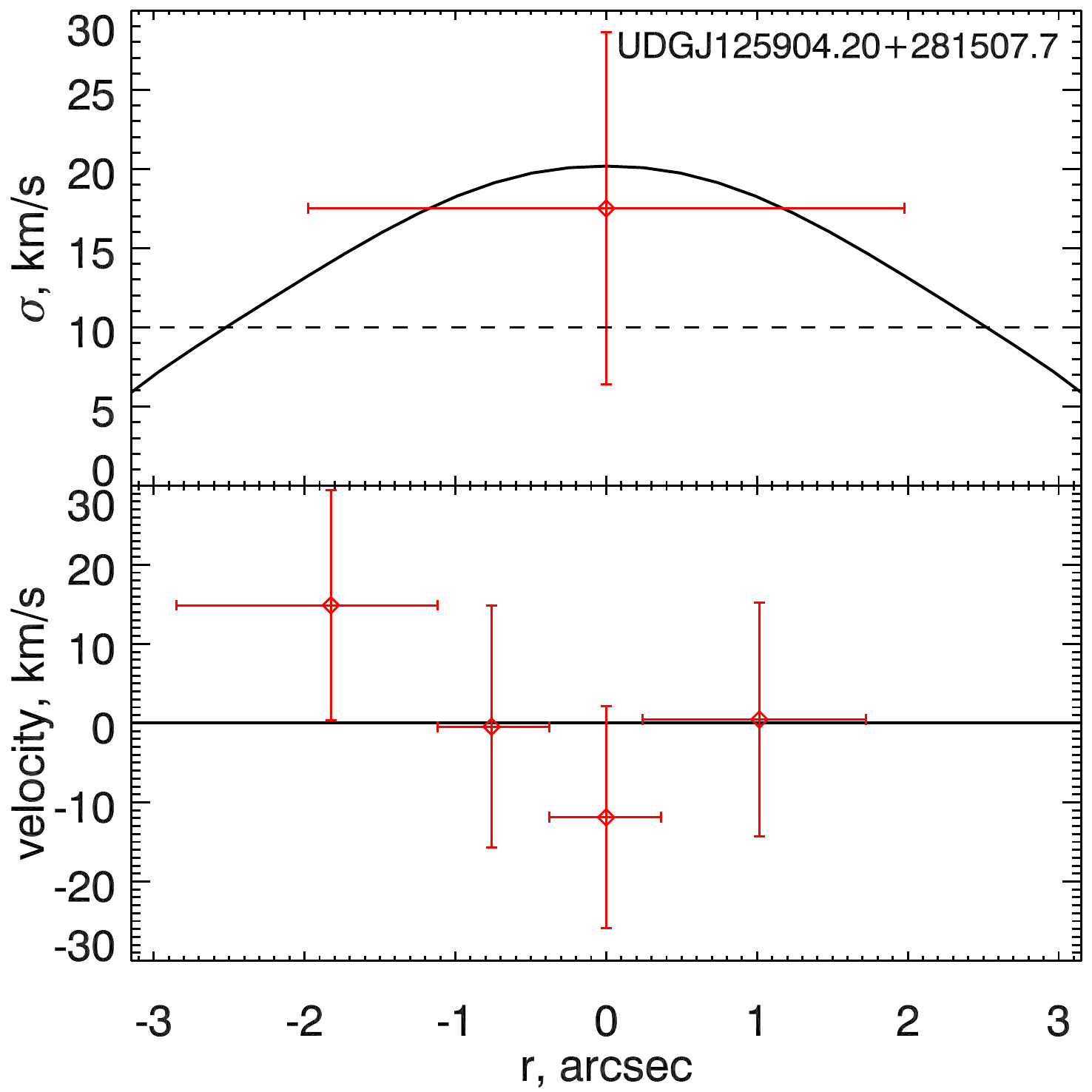} & 
\includegraphics[clip,trim={0cm 0 0.10cm 0.0cm}, width=0.41\hsize]{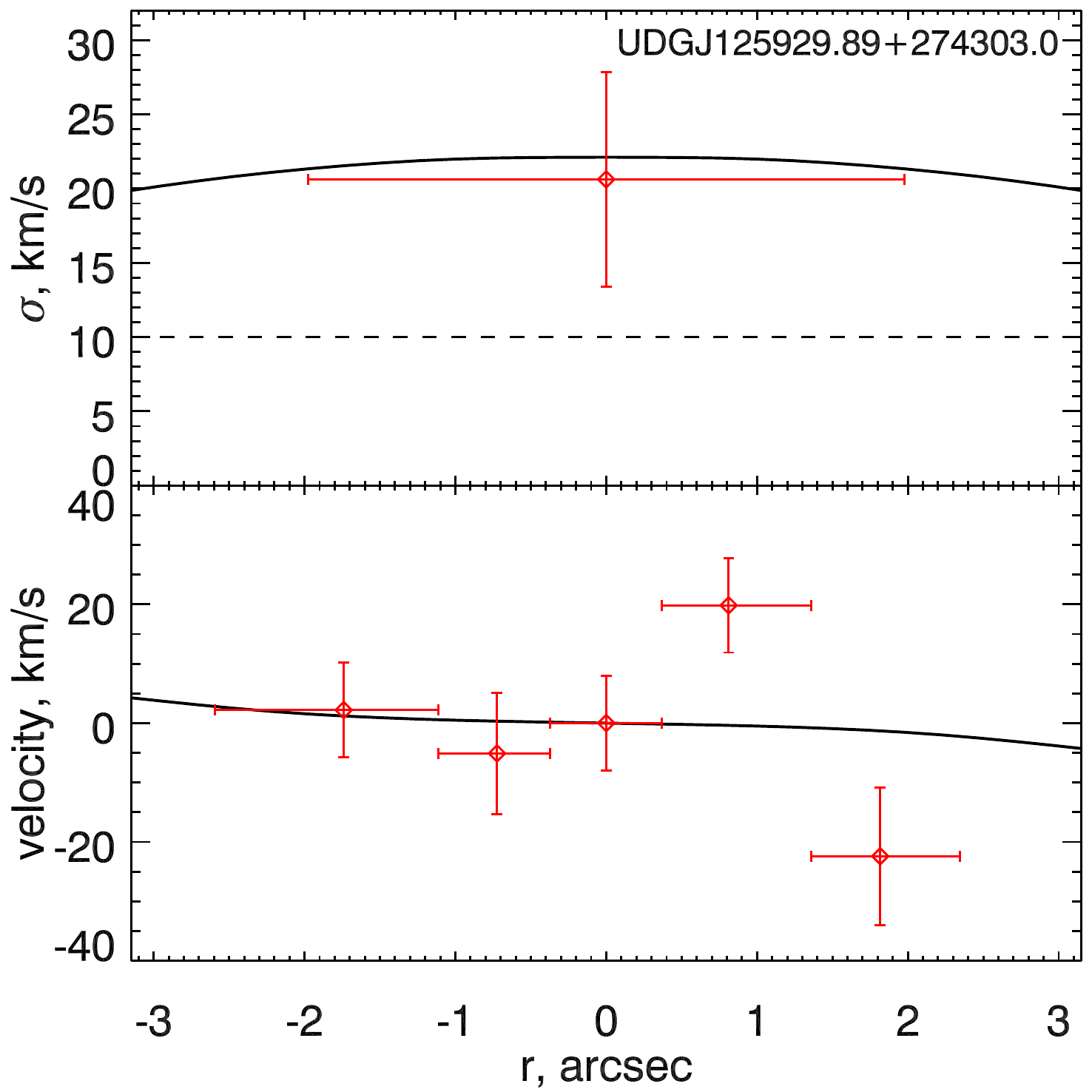} \\
\includegraphics[clip,trim={0.10cm 0 0.05cm 0.05cm}, width=0.41\hsize]{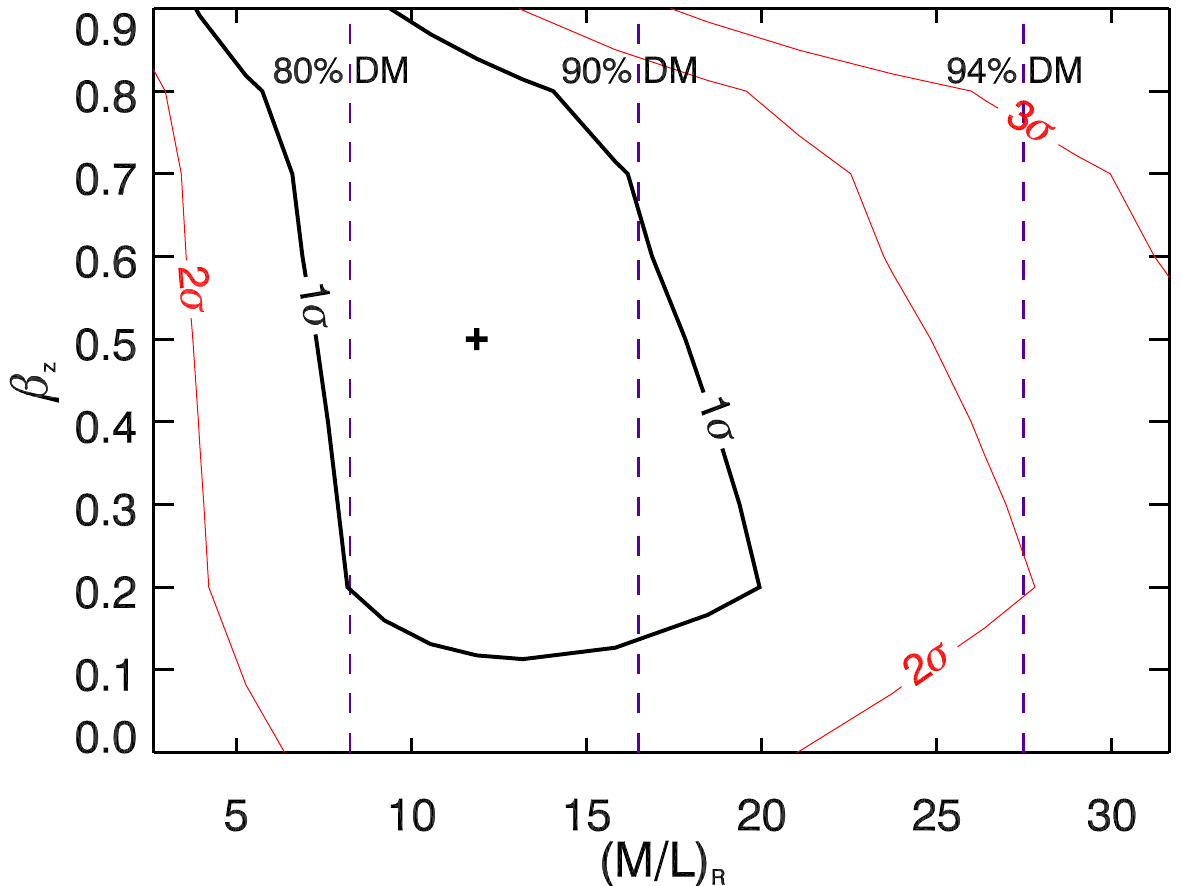} & 
\includegraphics[clip,trim={0.10cm 0 0.05cm 0.05cm}, width=0.41\hsize]{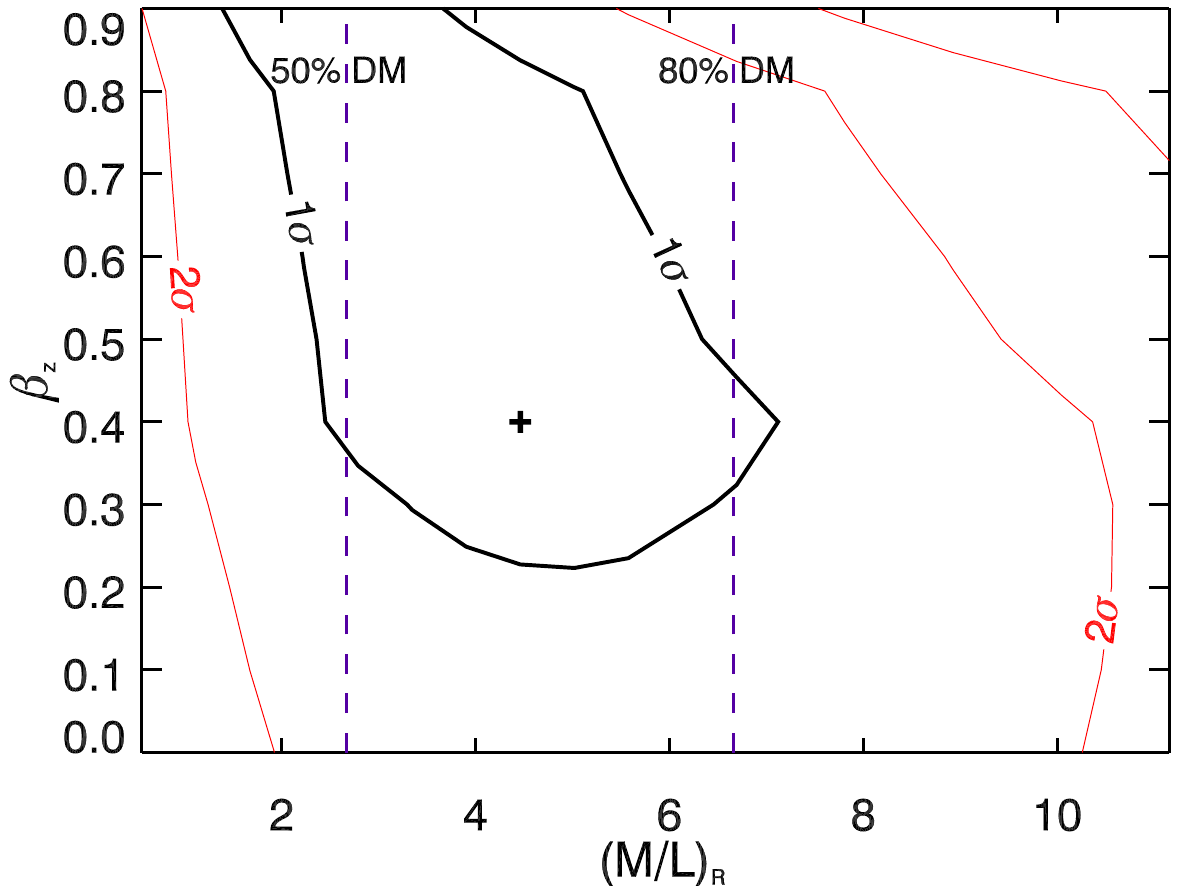} \\
\includegraphics[clip,trim={0.40cm 0.4cm 0.50cm 1.1cm}, width=0.425\hsize]{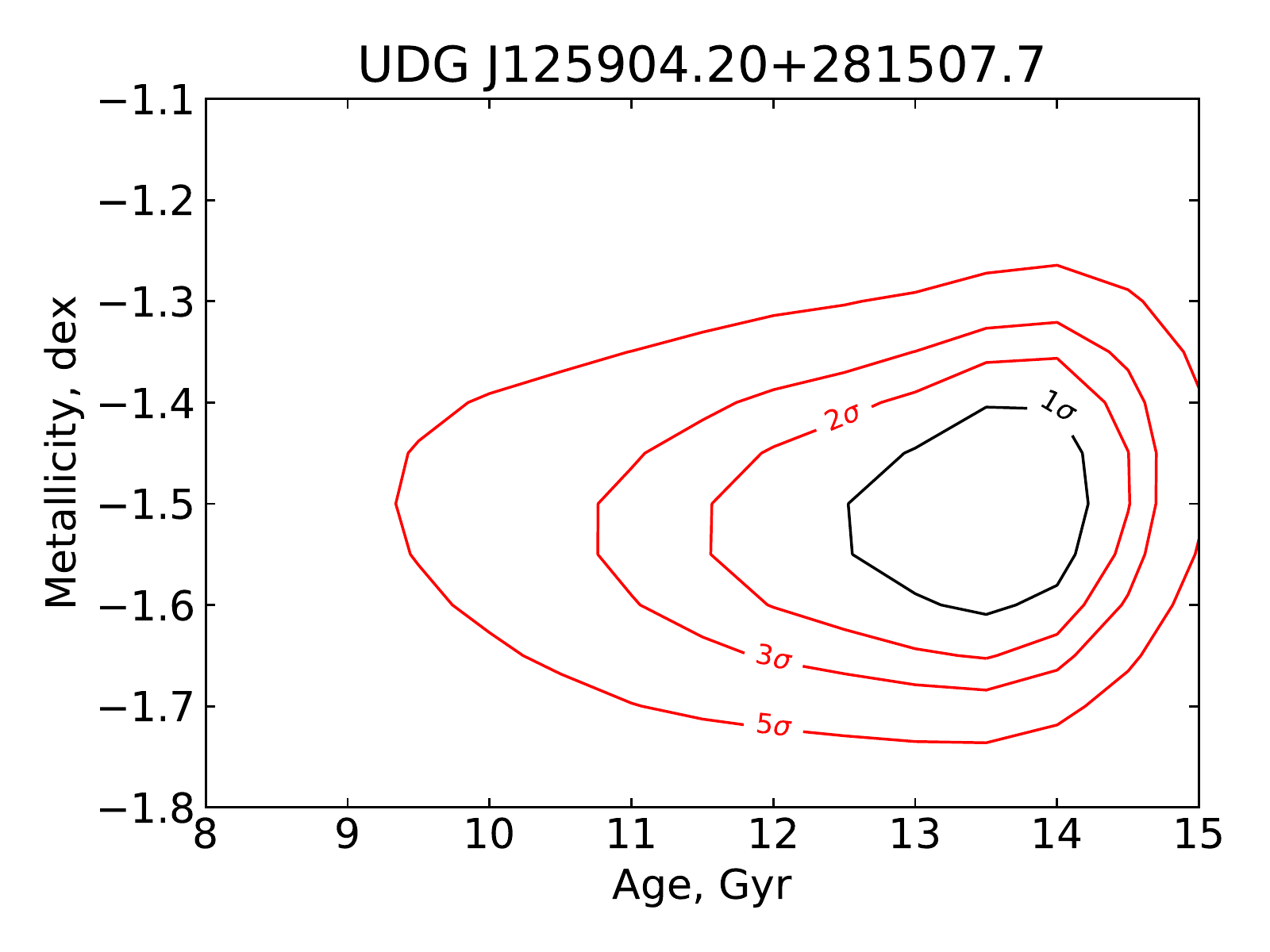} & 
\includegraphics[clip,trim={0.40cm 0.4cm 0.50cm 1.1cm}, width=0.425\hsize]{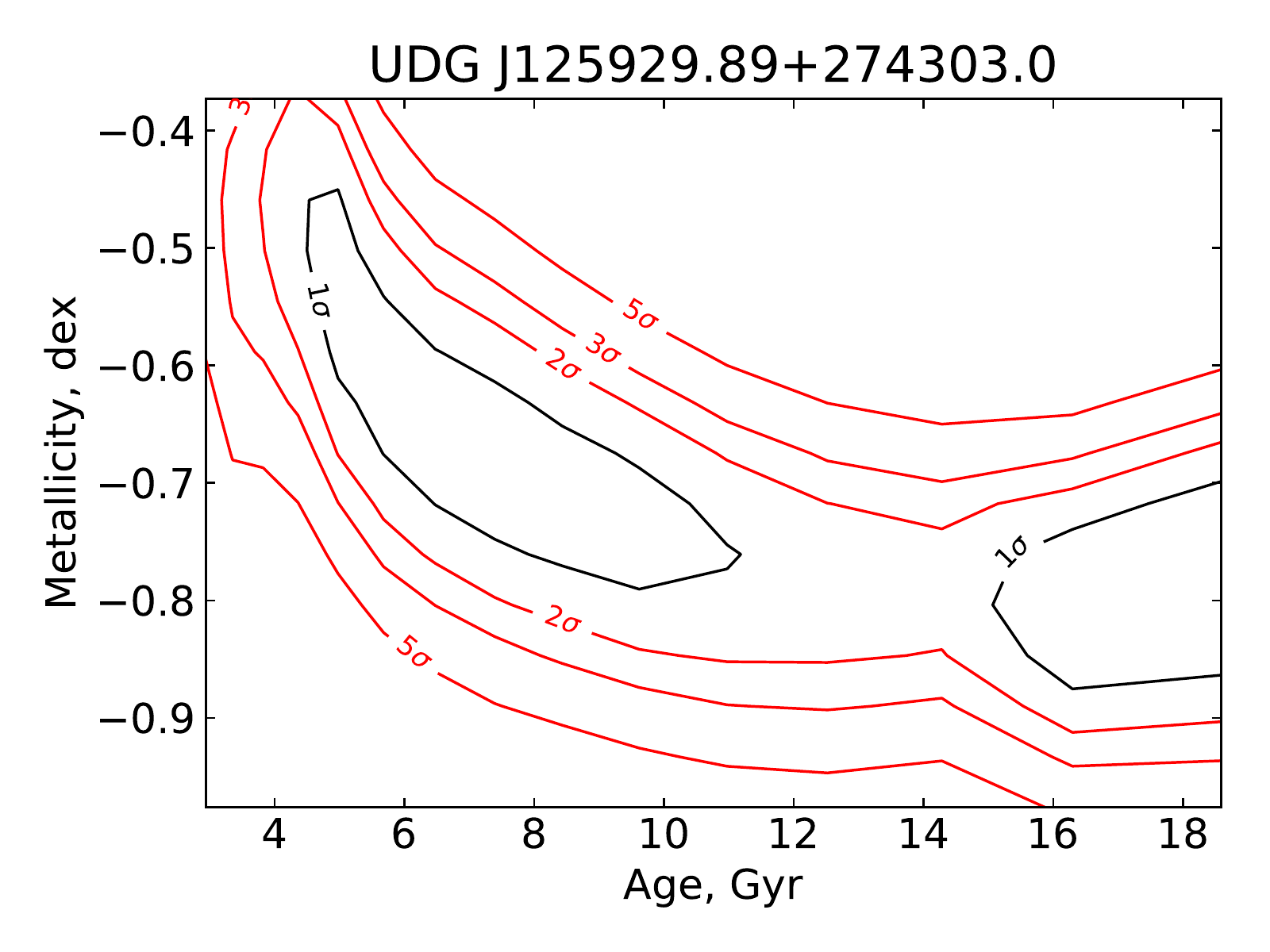} 
\end{tabular}
\caption{contd.}
\end{figure*}

\setcounter{figure}{1}
\begin{figure*}
\centering
\begin{tabular}{rr}
\includegraphics[angle=0,width=0.35\hsize]{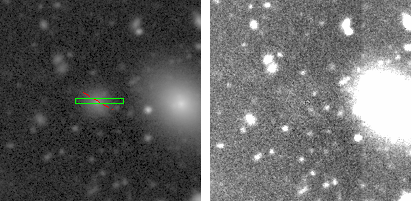} & 
\includegraphics[angle=0,width=0.35\hsize]{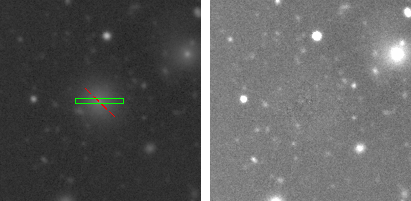} \\
\includegraphics[clip,trim={0cm 0 0.10cm 0.0cm}, width=0.41\hsize]{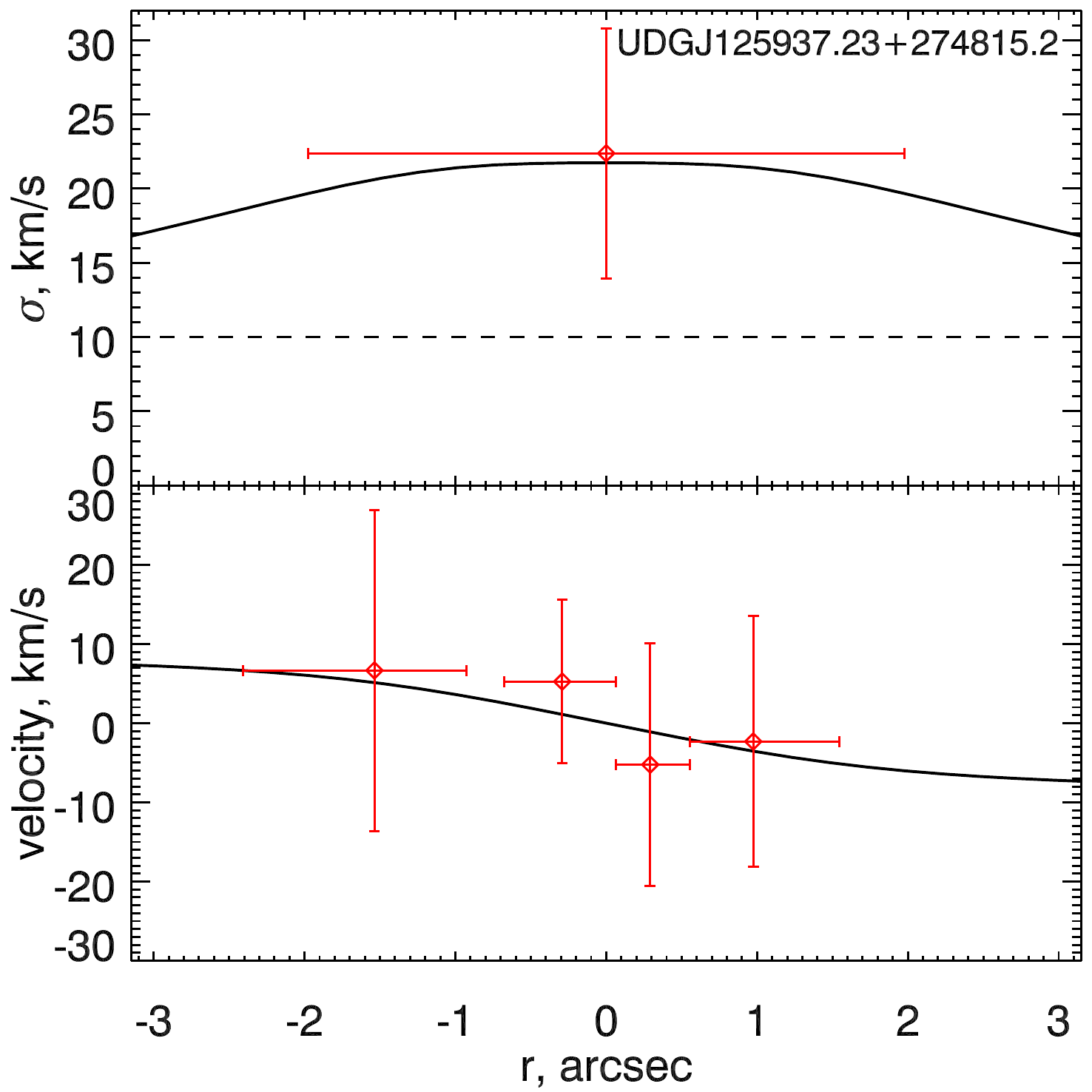} & 
\includegraphics[clip,trim={0cm 0 0.10cm 0.0cm}, width=0.41\hsize]{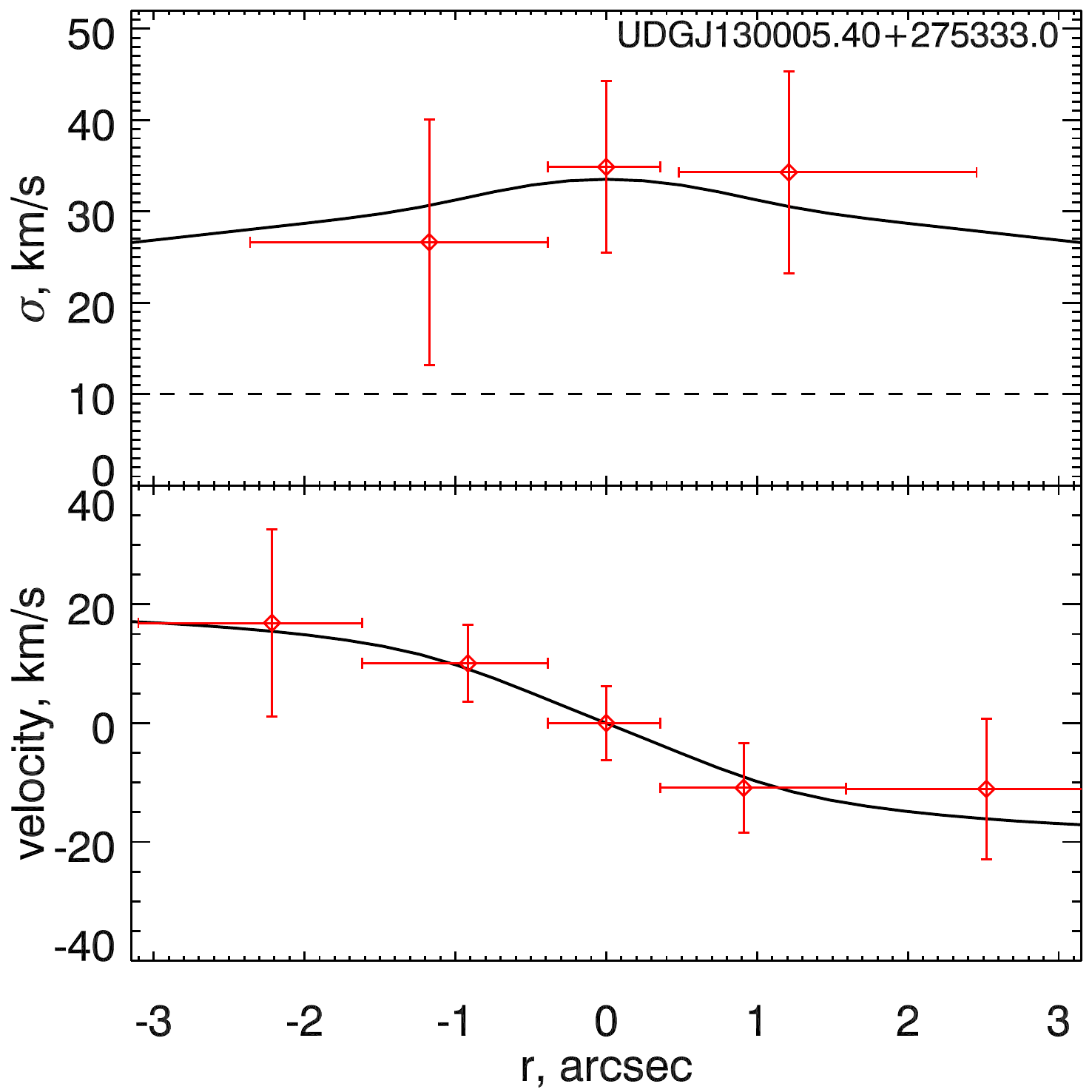} \\
\includegraphics[clip,trim={0.10cm 0 0.05cm 0.05cm}, width=0.41\hsize]{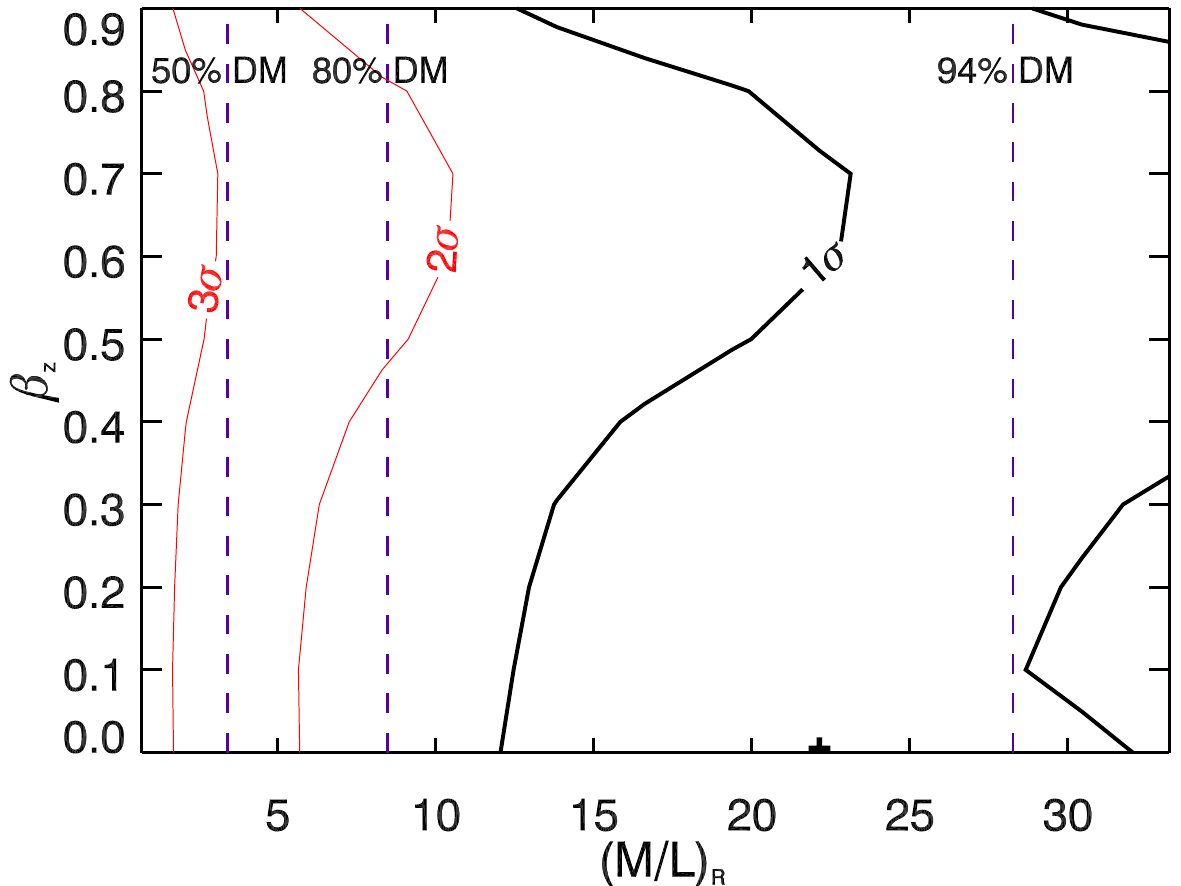} & 
\includegraphics[clip,trim={0.10cm 0 0.05cm 0.05cm}, width=0.41\hsize]{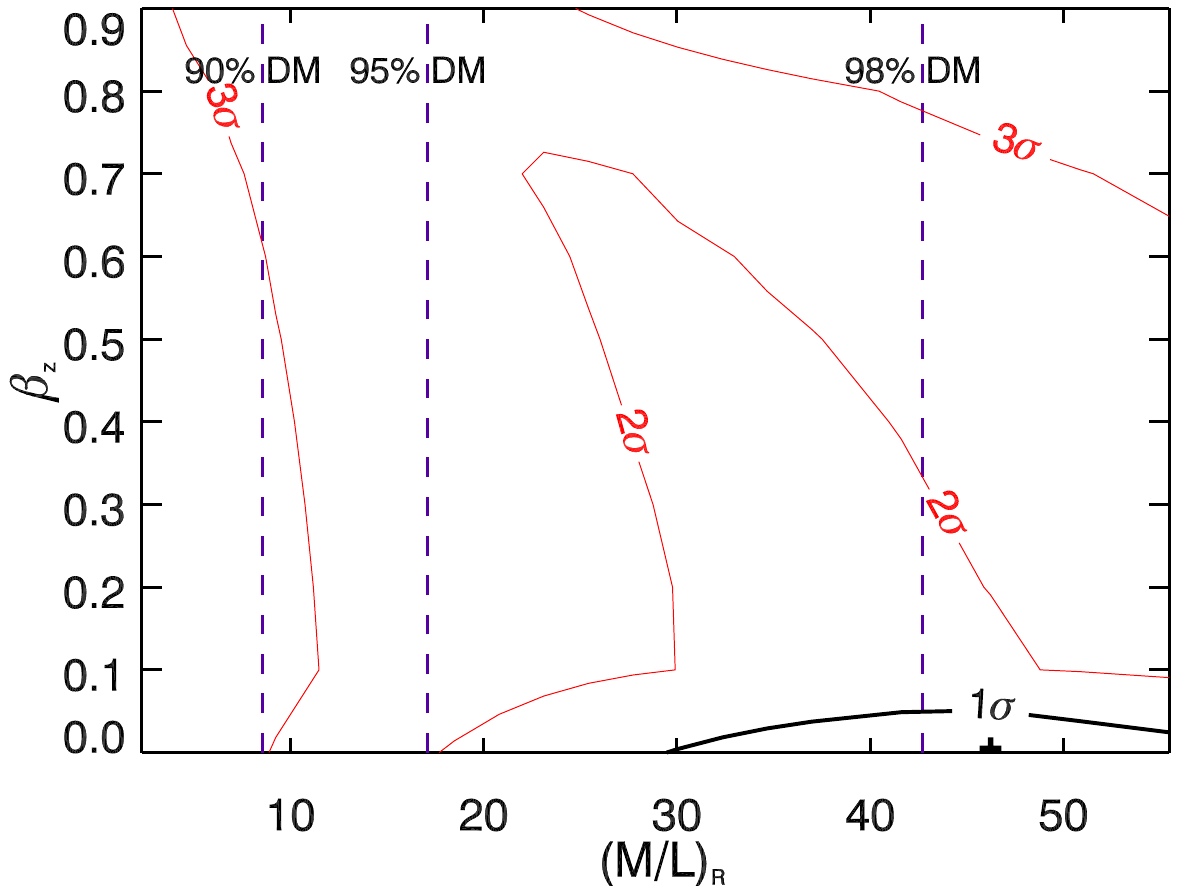} \\
\includegraphics[clip,trim={0.40cm 0.4cm 0.50cm 1.1cm}, width=0.425\hsize]{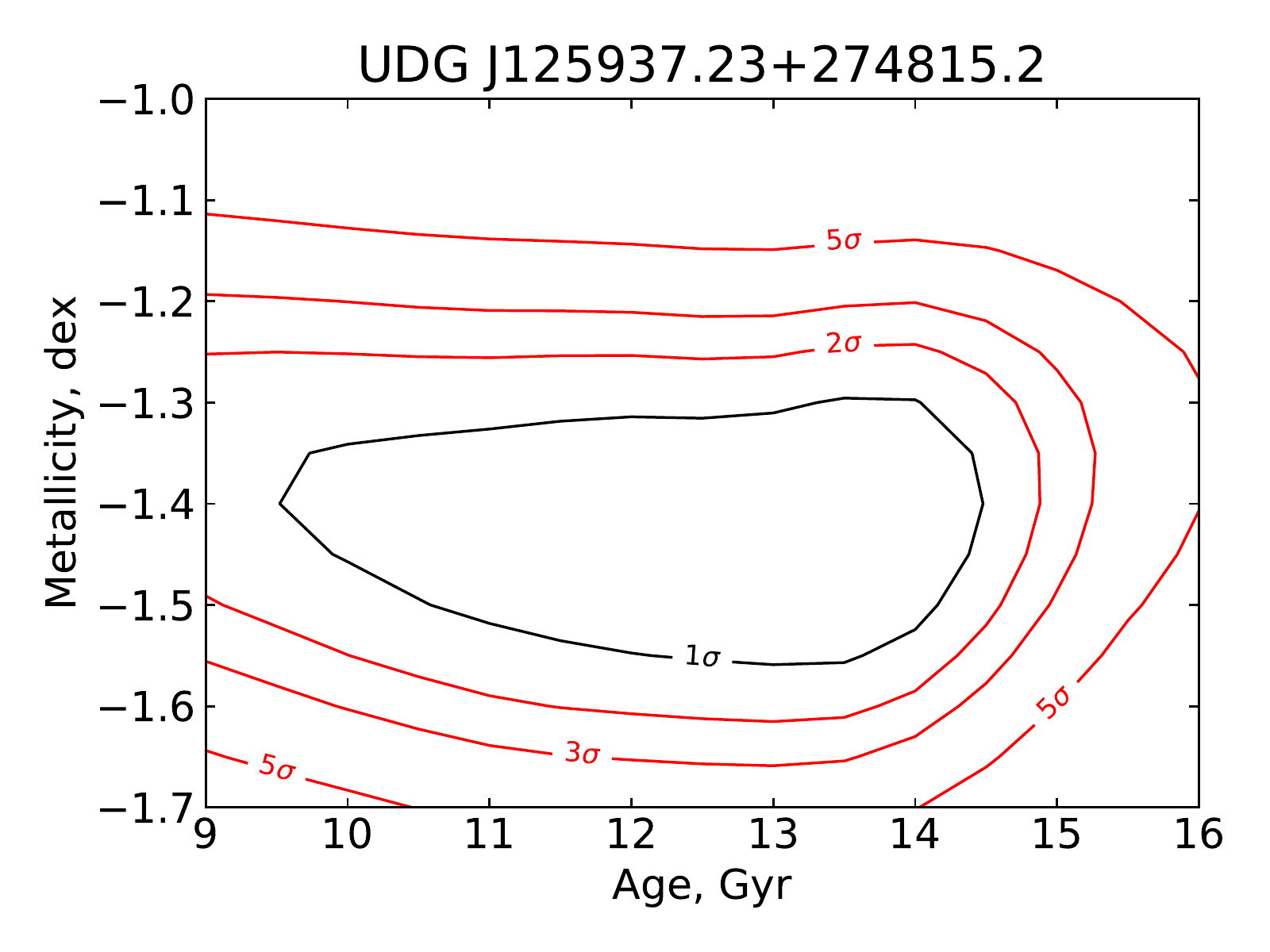} & 
\includegraphics[clip,trim={0.40cm 0.4cm 0.50cm 1.1cm}, width=0.425\hsize]{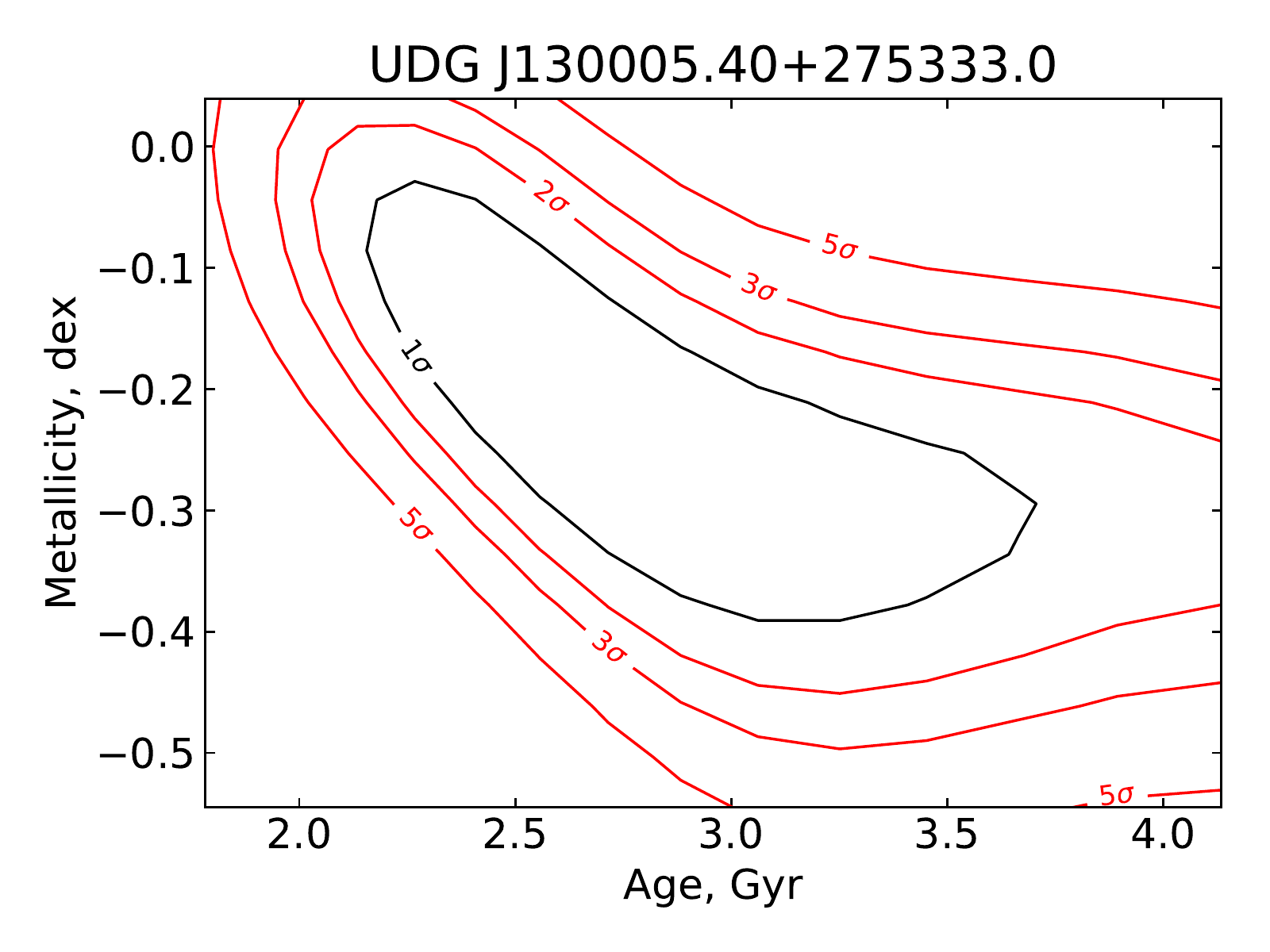}
\end{tabular}
\caption{contd.}
\end{figure*}

\setcounter{figure}{1}
\begin{figure*}
\centering
\begin{tabular}{rr}
\includegraphics[angle=0,width=0.35\hsize]{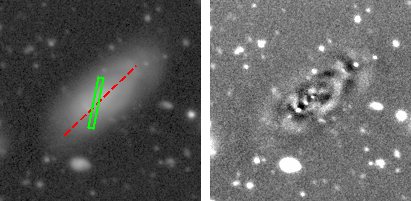} & 
\includegraphics[angle=0,width=0.35\hsize]{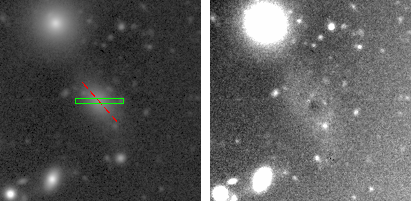} \\
\includegraphics[clip,trim={0cm 0 0.10cm 0.0cm}, width=0.41\hsize]{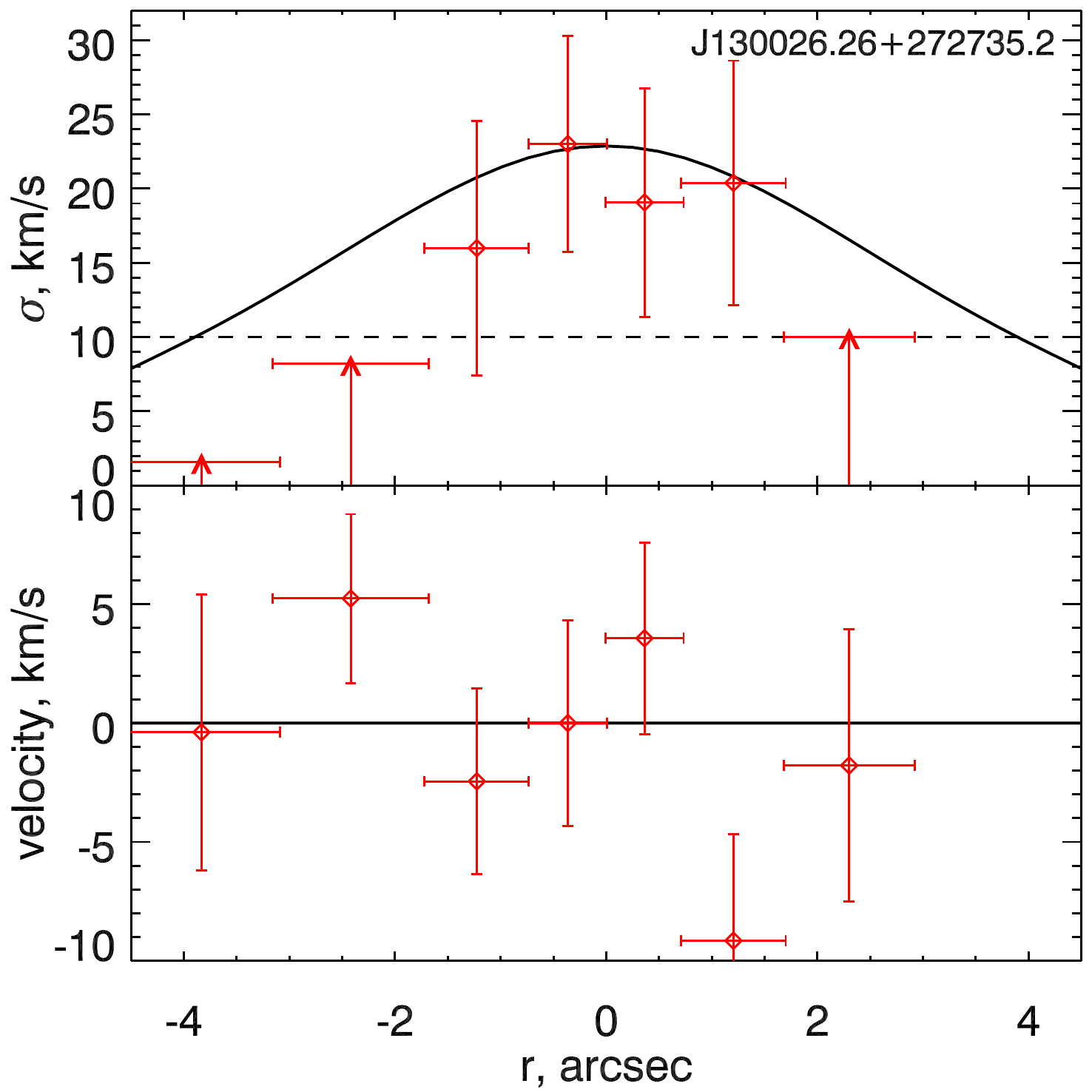} & 
\includegraphics[clip,trim={0cm 0 0.10cm 0.0cm}, width=0.41\hsize]{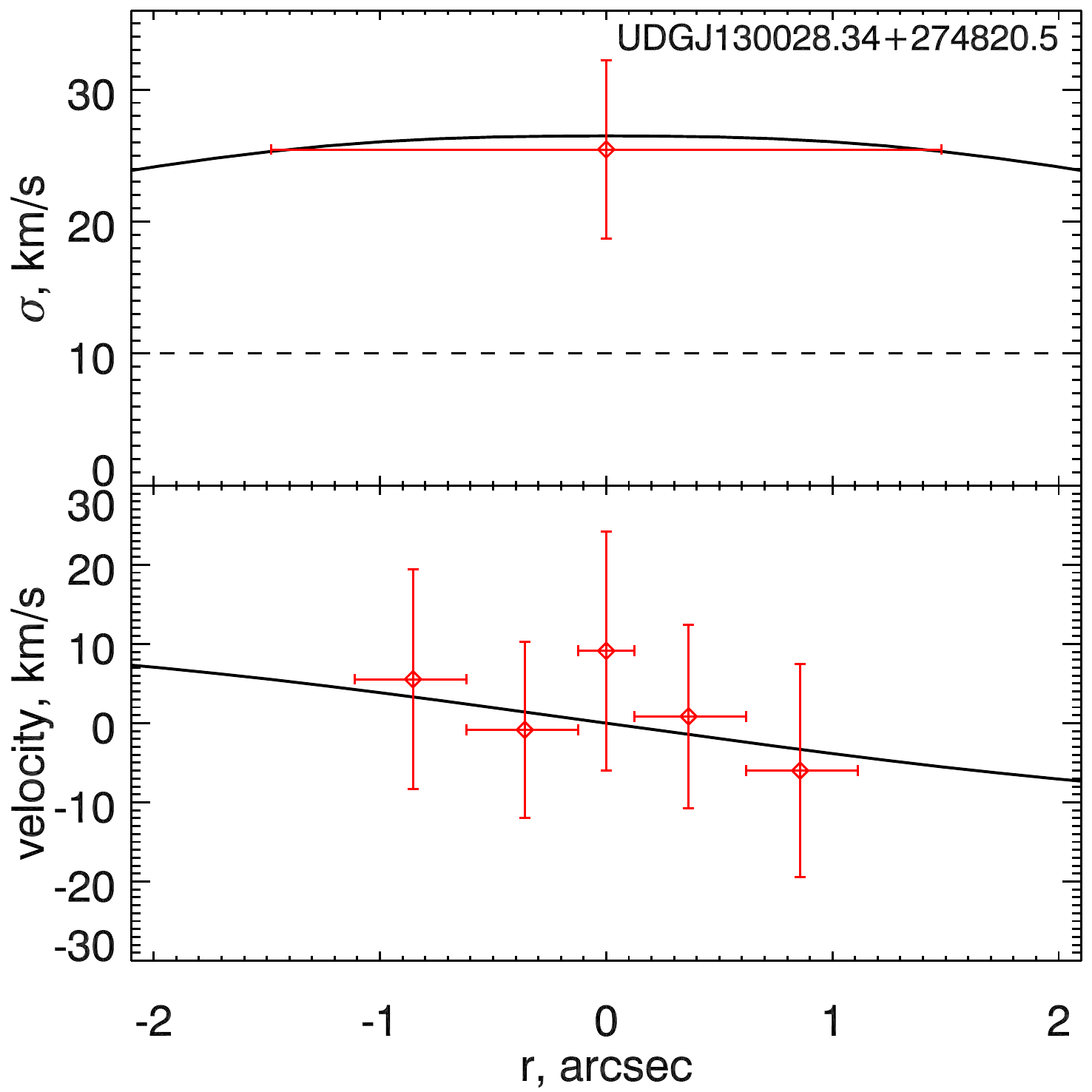} \\
\includegraphics[clip,trim={0.10cm 0 0.05cm 0.05cm}, width=0.41\hsize]{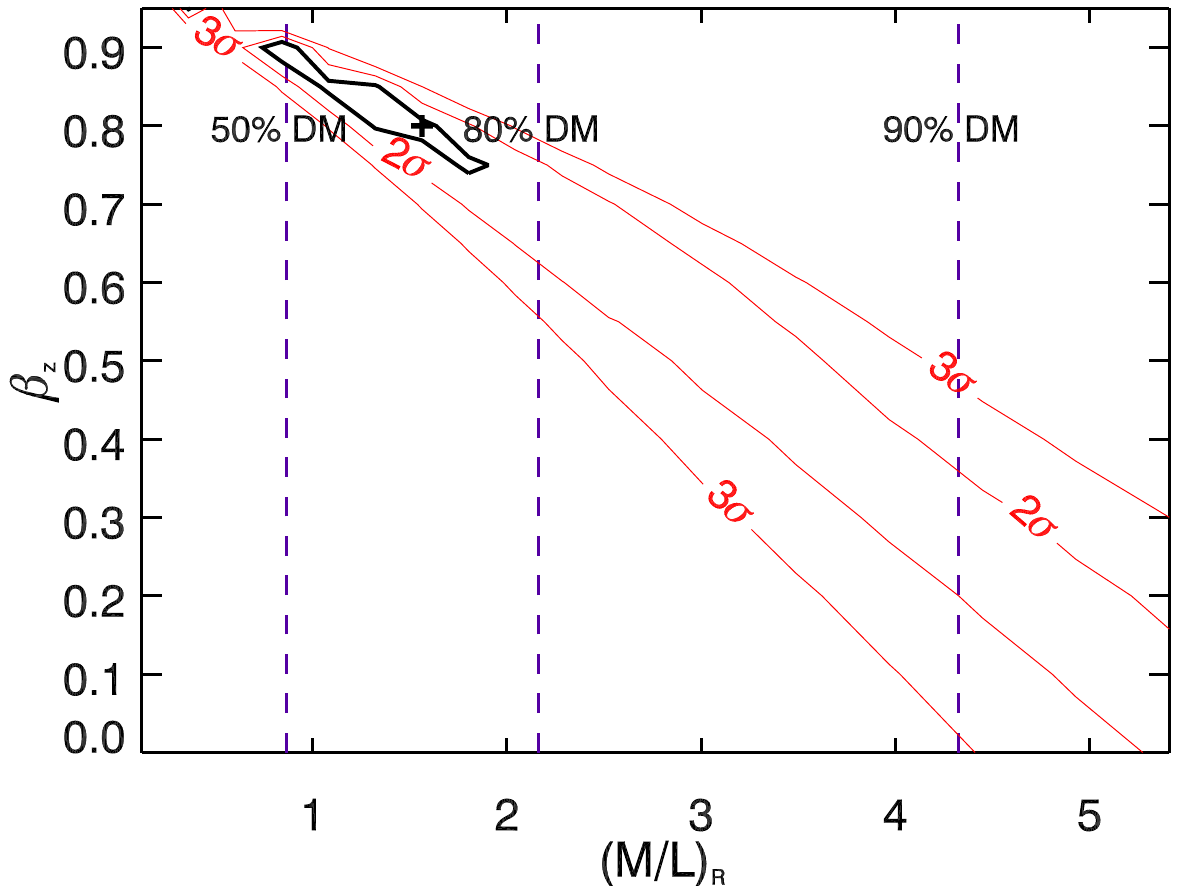} & 
\includegraphics[clip,trim={0.10cm 0 0.05cm 0.05cm}, width=0.41\hsize]{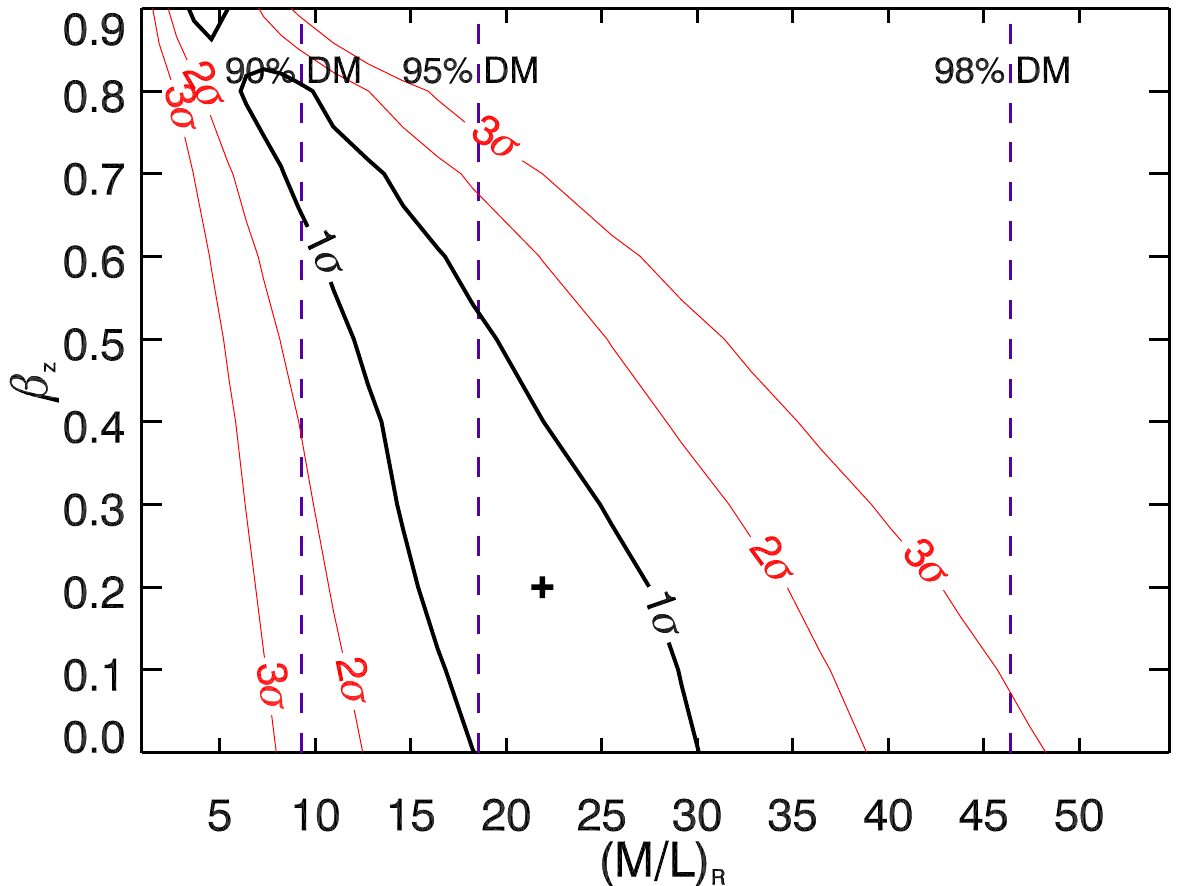} \\
\includegraphics[clip,trim={0.40cm 0.4cm 0.50cm 1.1cm}, width=0.425\hsize]{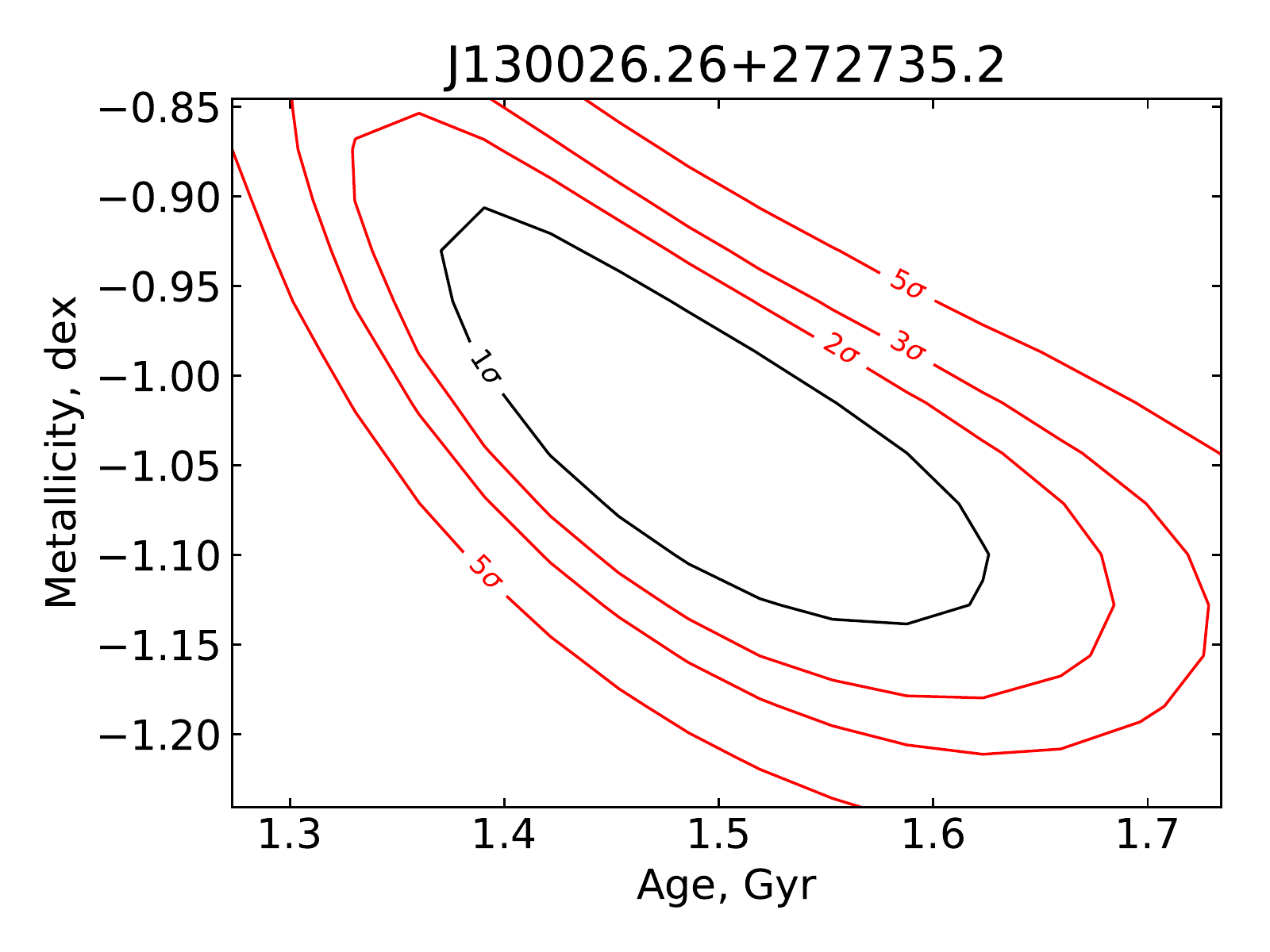} & 
\includegraphics[clip,trim={0.40cm 0.4cm 0.50cm 1.1cm}, width=0.425\hsize]{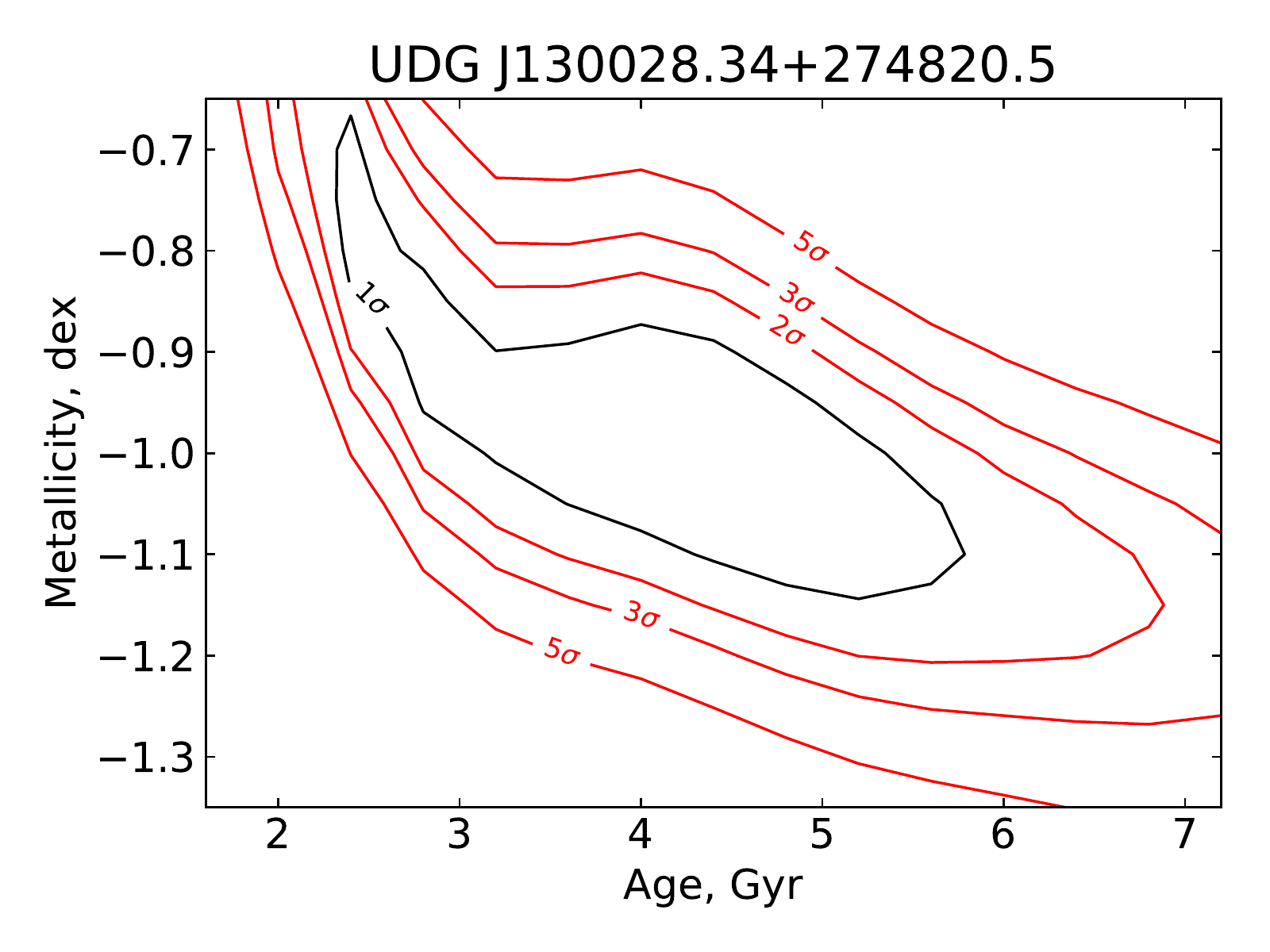}
\end{tabular}
\caption{contd.}
\end{figure*}

\setcounter{figure}{1}
\begin{figure}
\centering
\begin{tabular}{r}
\includegraphics[angle=0,width=0.75\hsize]{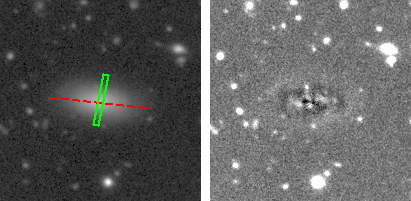}\\
\includegraphics[clip,trim={0cm 0 0.10cm 0.0cm}, width=0.88\hsize]{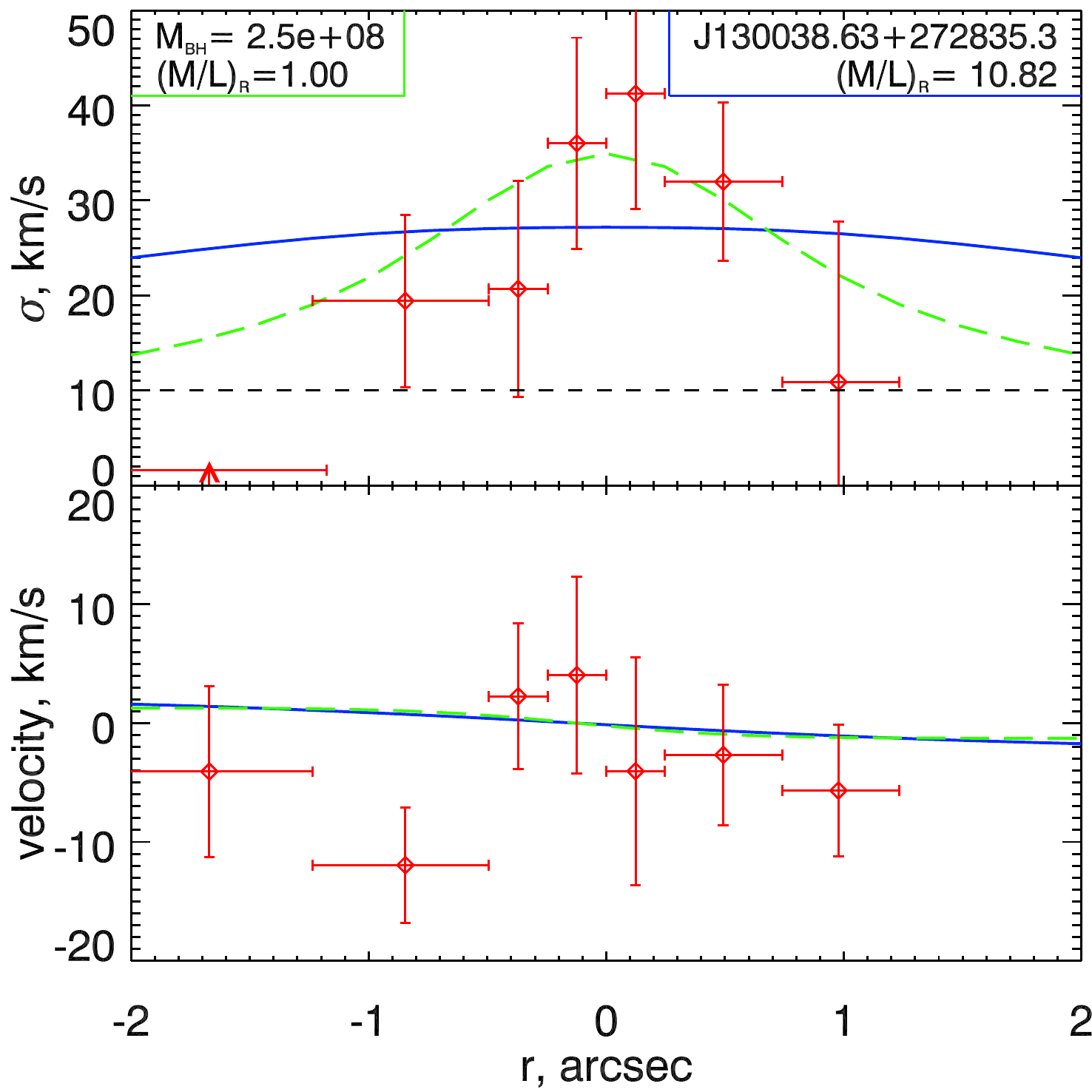}\\
\includegraphics[clip,trim={0.10cm 0 0.05cm 0.05cm}, width=0.88\hsize]{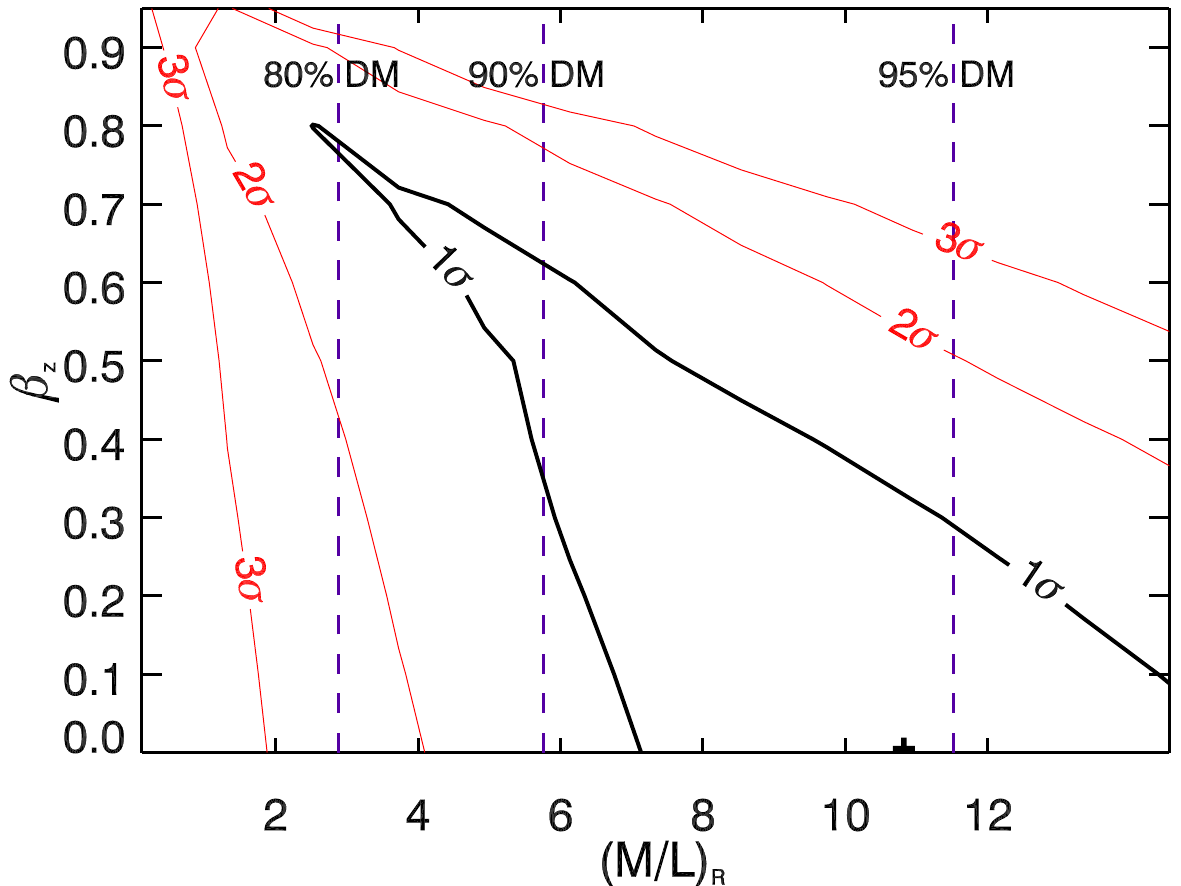}\\
\includegraphics[clip,trim={0.40cm 0.4cm 0.50cm 1.1cm}, width=0.93\hsize]{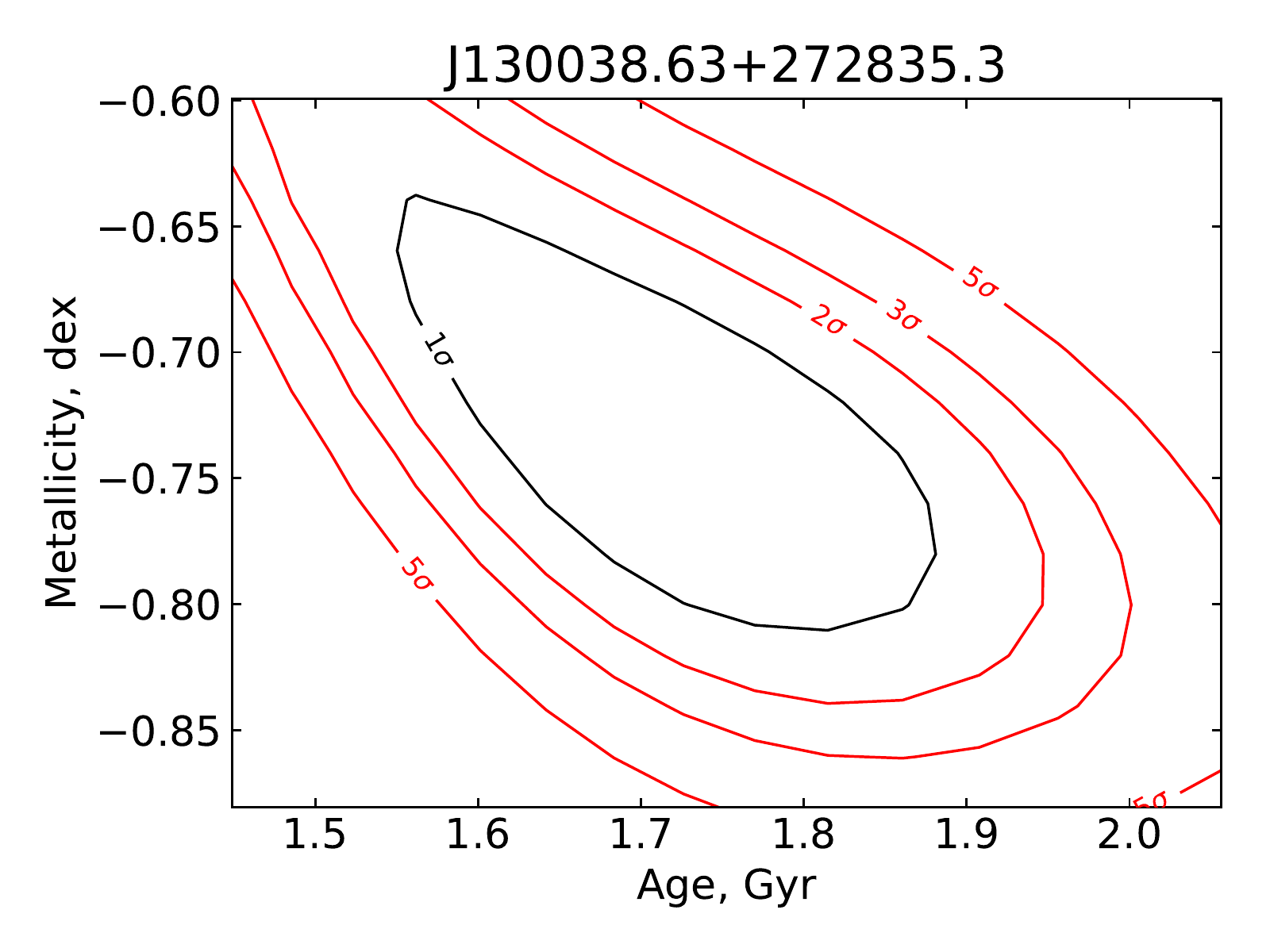}\\
\end{tabular}
\caption{contd. A physically improbable model with a central black hole for \emph{J130038.63+272835.3} is shown in green.}
\end{figure}

Deep Subaru Suprime-Cam images \citep{Koda15,yagi16} are available for the 9 UDGs. We extracted calibrated 60$^{\prime\prime}$ $\times$60$^{\prime\prime}$ cutout images from the Subaru archive  with {\sc topcat} \citep{2005ASPC..347...29T} implementing Virtual Observatory Simple Image Access. The images have a plate scale of 0.202$^{\prime\prime}$/pix with 0.65$^{\prime\prime}$ to  0.85$^{\prime\prime}$ FWHM seeing. We determined photometric \emph{AB} zero-points by matching the Subaru photometry of point sources to the DECaLS Data Release 7 catalog \citep{2018arXiv180408657D}. We estimated structural parameters for the 9 UDGs in the $V$ and $R$ bands using {\sc galfit v3} \citep{Peng10}. Four UDGs can be well described by a single-component S\'ersic model while the remaining five required an additional structural component. We used PSF models for foreground stars and unresolved background galaxies. Galaxy images and residuals from the best-fitting {\sc galfit} models are shown in the two top panels for every object in Fig.~\ref{udg_models}. The effective radii of galaxies range from 0.9~kpc to 3.8~kpc, which satisfy the \citet{Koda15} criterion ($r_e>0.7$~kpc).

We use the Jeans Anisotropic Modeling (JAM) approach \citep{Cappellari08} to estimate dynamical mass-to-light ratios and mass of dark matter in each galaxy. We follow a procedure described in \citet{2018MNRAS.477.4856A} without the black hole component. First, we convert the {\sc galfit} structural parameters into the Multiple Gaussian Expansion \citep{Cappellari02}, which we then use as input for dynamical modeling assuming that mass follows light. 
We then run the JAM over a parameter grid covering a wide range of $M/L_R$ ratios (0.5 to 32), anisotropy values $\beta_z$ (0 to 0.9), and  inclinations ($i$) from 90$^{\circ}$ (edge-on) down to the minimum allowed by the galaxy ellipticity (assuming an infinitely thin disk) in 5$^{\circ}$ steps. 
We compute $\chi^2$ statistics using the measured spatially resolved radial velocities and integrated velocity dispersions at every point of the parameter grid. We consider values of $\sigma<$10~km~s$^{-1}$ to be lower limits in the calculation of $\chi^2$ for \emph{J130026.26+272735.2} and \emph{J130038.63+272835.3} where we have spatially resolved $\sigma$ profiles. Finally, we construct a 3-dimensional $\chi^2$ map for each galaxy and determine the best-fitting values of $M/L$, $\beta_z$ and $i$. 

We also estimate the $M/L_R$ ratios using \citet{Wolf+10} technique to allow comparison with earlier work. The Wolf estimates exceed the JAM $M/L_R$ by 20--800\%\ depending on the galaxy. This discrepancy arises from the differing assumptions in the two approaches. \citet{Wolf+10} estimates are calculated using spherical Jeans equations, while JAM uses axisymmetric equations. \citet{Wolf+10} estimates are not properly applied to oblate or disky galaxies, which are present in our sample. The greatest difference between the two $M/L$ estimators (almost an order of magnitude) is found in the most flattened of the 9 galaxies (q=0.5), and the lowest (20\%) difference in $M/L$ is for a relatively round galaxy (q=0.82). We conclude that the JAM technique is preferred for our galaxy sample.

However, even the JAM technique does not allow us to accurately determine all dynamical parameters with an integrated $\sigma$ measurement and 3 to 6 velocity measurements along a slit that is not aligned with the galaxy major axis. The 1-$\sigma$ confidence contours in $\beta_z - M/L$ parameter space are large in 6 out of 9 galaxies. However, we can still constrain the dynamical $M/L$ ratios and the dark matter content of the 9 UDGs.

\section{Results and Discussion}

\subsection{Internal properties of UDGs}

In Table~\ref{tab:udgprop} we present internal kinematics, stellar population properties and stellar mass-to-light ratios of 9 UDGs extracted from integrated spectra, structural properties derived from the analysis of Subaru $R$-band images, and dynamical mass-to-light ratios, orbital anisotropy parameters $\beta_z$ and inclinations to the line of sight obtained from JAM. We convert age and metallicities into stellar $M/L$ ratios using {\sc pegase.hr} model predictions for SSP models assuming a \citet{Kroupa02} stellar initial mass function.

\emph{J125848.94+281037.1} is a flattened galaxy ($b/a=0.65$) observed along the major axis. This is the most metal-poor object in our sample ([Fe/H]$=-$1.8~dex). There is a hint of major axis rotation from the three data points along the slit. We are able to constrain both orbital anisotropy (close to 0) and the dark matter content ($\sim90$\%).

\emph{J125904.06+281422.4} is a flattened galaxy ($b/a=0.56$) observed almost along the major axis. There is a hint of rotation ($<10$~km/s) in the outskirts and a flat velocity profile in the central region that drive the dynamical model towards higher anisotropy values with a dark matter content around $\sim90$\%\ by mass.

\emph{J125904.20+281507.7} is an approximately round ($b/a=0.82$), metal-poor ([Fe/H]$=-$1.5~dex) galaxy observed along the major axis. The radial velocity profile does not show significant rotation and can be fit with a moderate mass-to-light ratio with anisotropic orbits ($\beta_z=0.5$), assuming axial symmetry.

\emph{J125929.89+274303.0} is a strongly flattened galaxy ($b/a=0.50$) observed about 30~deg off the minor axis. We see no radial velocity gradient and dynamical modelling suggests highly anisotropic orbits. Its stars are more metal rich ($-0.6$~dex) than most of the other galaxies in our sample, and the age of the stellar population is around 6--8~Gyr. This galaxy is included in the sample of \citet{FerreMateu+18} as \emph{Yagi~275}; they report a somewhat higher metallicity value ($-0.37$~dex). 

\emph{J125937.23+274815.2} is a nearly round galaxy ($b/a=0.87$) observed 18~deg off the major axis with the smallest effective radius (0.9 kpc) in our sample. We are unable to constrain the orbital anisotropy. This galaxy is not physically connected to the brighter \emph{GMP~3324}  visible to the right as their radial velocities differ by almost 1000~km/s.

\emph{J130005.40+275333.0} is quite round ($b/a=0.9$) with a distinct stellar nucleus unresolved both on Subaru and archival HST ACS images. The nuclear stellar population has the same age (3~Gyr) as the rest of the galaxy but may be more metal-rich ([Fe/H]$=-0.08\pm0.18$~dex vs $-0.30\pm0.22$~dex). The rotation ($20\pm5$~km/s at 0.5~kpc from the center) cannot be fit by axisymmetric models with adequate $M/L$ ratios. We suspect a triaxial system similar to some dwarf \citep[\emph{VCC~1545},][]{Chilingarian09} and giant \citep[\emph{NGC~4365},][]{2001ApJ...548L..33D} elliptical galaxies. A tentative detection of minor axis rotation in the UDG \emph{NGC~1052-DF2} \citep{2018arXiv181207345E} is consistent with a triaxial shape, but velocity measurements at higher spectral resolution \citep{Danieli+19} suggest a flat velocity field.

\emph{J130026.26+272735.2} is the most spatially extended galaxy in our sample. Its slightly asymmetric shape explains the offset of the coordinates reported in \citet{Adami08} and used in our mask design so that the slit landed 1$^{\prime\prime}$ off the galaxy nucleus. We find a relatively young stellar population (1.5~Gyr) but no current star formation evidenced by the absence of emission lines. The absence of rotation with strong flattening ($b/a=0.44$) suggest that stars in its disk have highly anisotropic orbits.  This is best case in our sample where the $\beta_z - M/L$ degeneracy is broken by dynamical modeling.

\emph{J130028.34+274820.5}. Dynamical modelling yields a very high $M/L$ ratio and isotropic stellar orbits similar to \emph{J125937.23+274815.2}.  The data are consistent with a weak velocity gradient along the slit placed about 50~deg off the major axis. S-shaped image fitting residuals and the weak rotation may indicate tidal interaction with a nearby brighter galaxy (\emph{J130028.98+274836.2}) visible in the top left corner of the image cutout. However no radial velocity measurement for \emph{J130028.98+274836.2} is available so the two galaxies may be physically separate.

\emph{J130038.63+272835.3} has a relatively young stellar population ($2$~Gyr) and a strongly flattened shape ($b/a=0.42$). The central bump in the velocity dispersion profile ($40\pm7$~km/s compared to $15\pm10$~km/s outside the peak) cannot be explained with standard dynamical modelling. If the bump is caused by a central black hole (dashed green line in Fig.~\ref{udg_models}) a mass of $2.5 \cdot 10^8 M_{\odot}$ would be required, $\sim$3 times the total stellar mass of the galaxy ($\sim7.5 \cdot 10^7 M_{\odot}$). Deep integral field unit observations would clarify the nature of this galaxy.

\begin{deluxetable*}{lD{,}{\pm}{-1} D{,}{\pm}{-1}D{,}{\pm}{-1}D{,}{\pm}{-1} D{,}{\pm}{-1}D{,}{\pm}{-1}D{,}{\pm}{-1}ccccc} 
\setlength{\tabcolsep}{2pt}
\small
\tablecaption{UDGs and their properties \label{tab:udgprop}}
\tablecolumns{14}
\tablenum{1}
\tablewidth{0pt}
\tablehead{
\colhead{IAU name} & 
\colhead{$v_{\mathrm{sys}}$} & 
\colhead{$\sigma_{\mathrm{eff}}$ }& 
\colhead{Age} & 
\colhead{[Fe/H]} & 
\colhead{(M/L)$_{R*}$} & 
\colhead{(M/L)$_{R,\mathrm{d}}$} & 
\colhead{$\beta_z$} & 
\colhead{$i$} & 
\colhead{r$_e$} & 
\colhead{$M_R$} & 
\colhead{$\langle \mu_{e,R} \rangle$} \\
\colhead{} &
\colhead{km/s} &
\colhead{km/s} &
\colhead{Gyr} &
\colhead{dex} &
\colhead{(M/L)$_{\odot}$} & \colhead{(M/L)$_{\odot}$} &
\colhead{} & \colhead{deg} & \colhead{kpc} & \colhead{mag} & \colhead{mag/arcsec$^{2}$}
}

\startdata
J125846.94+281037.1 & 5486,8 & 18,13 &  12.5,1.8 & -1.78,0.15 & 1.32,0.16 & 16,4 &  0.1,0.1 & 70 &  1.3 & -14.39 & 24.65 \\
J125904.06+281422.4 & 6436,5 & 25,7 & 12.6,2.2 & -1.06,0.12 & 1.48,0.40 & 22,8 & 0.6,0.2 & 75 & 1.3 & -14.59 & 24.45 \\ 
J125904.20+281507.7 & 6910,6 & 17,10 & 13.5,1.2 & -1.41,0.17 & 1.41,0.11 & 12,5 & 0.5,0.3 & 60 & 1.2 & -14.46 & 24.46 \\ 
J125929.89+274303.0 & 4928,4 & 21,7 & 7.4,1.7 & -0.67,0.17 & 1.14,0.25 & 4.5,2.5 & 0.4,0.2 & 75 & 2.1 & -15.38 & 24.67 \\ 
J125937.23+274815.2 & 7305,5 & 22,8 & 12.4,2.8 & -1.26,0.13 & 1.45,0.12 & 22,10 & 0.0,1 & 75 & 0.9 & -14.58 & 23.76 \\ 
J130005.40+275333.0\tablenotemark{*} & 6284,4 & 37,6 & 2.7,0.3 & -0.25,0.17 & 0.73,0.11 & 47,17 & 0.0,0.1 & 50 & 1.4 & -15.33 & 23.80 \\ 
J130026.26+272735.2 & 6939,2 & 19,5 & 1.5,0.1 & -1.04,0.11 & 0.37,0.02 & 2.0,0.6 & 0.8,0.1 & 70 & 3.7 & -16.97 & 24.33 \\ 
J130028.34+274820.5\tablenotemark{*} & 6471,4 & 23,7 & 4.7,0.8 & -1.01,0.18 & 0.79,0.12 & 22,6 & 0.2,0.3 & 60 & 1.3 & -14.87 & 24.18 \\ 
J130038.63+272835.3\tablenotemark{*} & 7937,3 & 27,5 & 1.7,0.1 & -0.74,0.08 & 0.41,0.03 & 11,4 & 0.0,0.4 & 72 & 1.9 & -16.03 & 23.80 \\
\enddata

\tablecomments{The columns are: IAU name, systemic radial velocity, integrated stellar velocity dispersion, mean age and metallicity of stellar population and stellar $M/L$ ratio derived with the full spectrum fitting using SSP models, dynamical M/L, orbital anisotropy parameter, and inclination to the line-of-sight obtained with JAM,  half-light radius of a galaxy, $R$-band \emph{AB} absolute magnitude, and effective surface brightness value de-projected using measured ellipticity of a galaxy image. All photometry is corrected for foreground extinction \citep{SF11}, surface brightness is also corrected for cosmological dimming. The $k$-corrections are $<0.01$~mag \citep{CMZ10}.}
\tablenotetext{*}{JAM is likely inapplicable to these objects (see the text for details).}
\end{deluxetable*}

\subsection{Evolutionary link with dwarf early-type galaxies}

We estimate the dark matter fraction within each galaxy's half-light radius by comparing stellar and dynamical $M/L$ ratios. Dashed purple vertical lines mark dark fractions in Fig.~\ref{udg_models} in the  panels showing confidence levels in $\beta_z - M/L$ space. For the two objects with high orbital anisotropy, \emph{J125929.89+274303.0} and \emph{J130026.26+272735.2}, the dark matter fraction estimate is 50~\%\ to 80~\%\, very similar to what was found in the three dE satellites of the Andromeda galaxy \citep{2006MNRAS.369.1321D}.  The Andromeda satellites also exhibit weak (\emph{NGC~205}) or no (\emph{NGC~147} and \emph{NGC~185}) rotation and high orbital anisotropy. The two galaxies with signs of major axis rotation, \emph{J125904.06+281422.4} and \emph{J125904.20+281507.7} likely have somewhat higher dark matter content, around 90~\% but further, deeper observations would be required for confirmation. More massive cluster dE/dS0s tend to have somewhat lower dark matter fractions within $r_e$ typically 30--50~\%\ \citep{Forbes+11}. 

The two galaxies with younger stellar populations, \emph{J130026.26+272735.2} and \emph{J130038.63+272835.3} likely entered the central region of the Coma cluster quite recently with gas-rich disks that were stripped by the ram pressure of the hot intracluster gas \citep{1972ApJ...176....1G}.

\begin{figure}
    \centering
    \includegraphics[width=0.95\hsize]{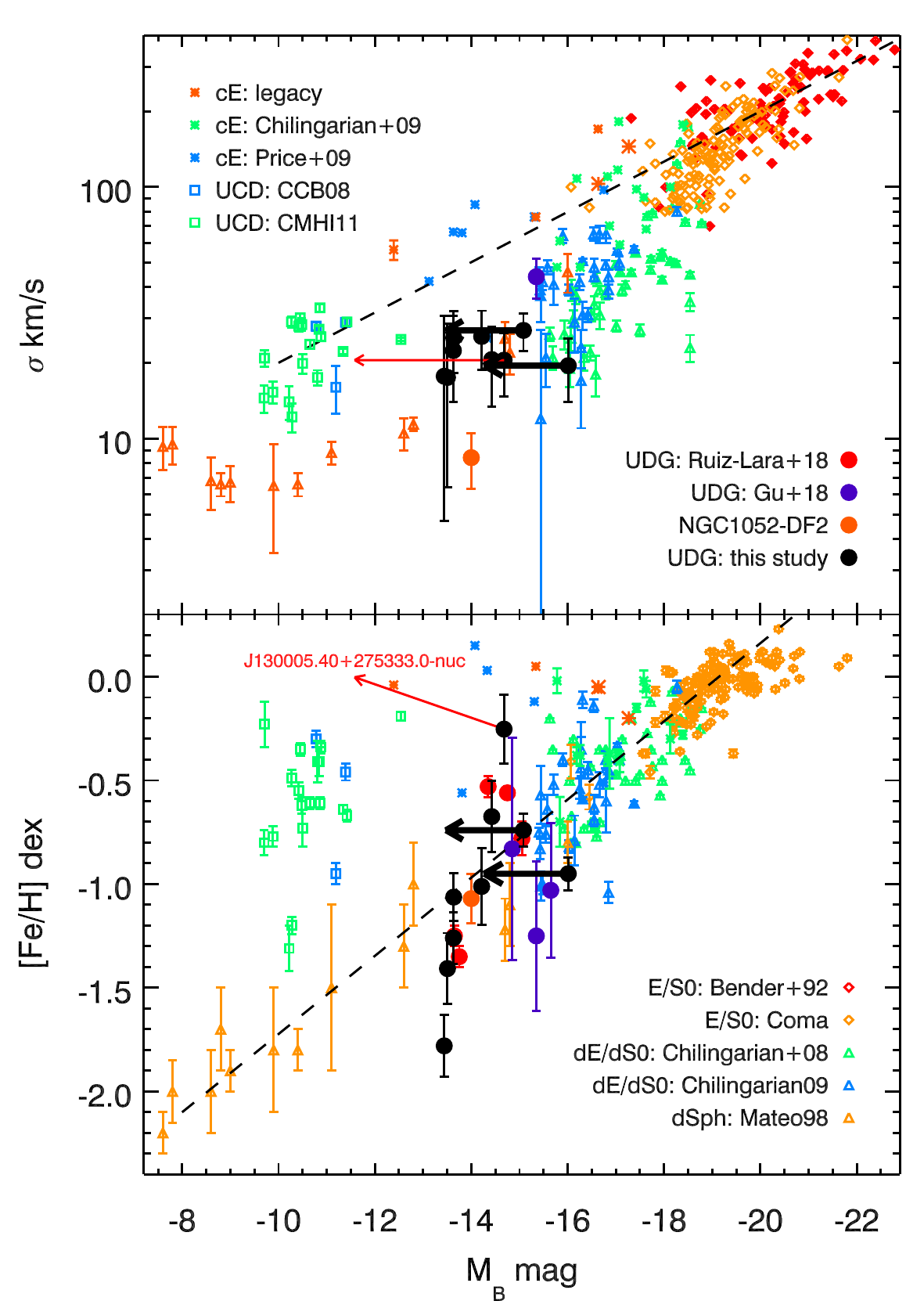}
    \caption{Faber--Jackson (top) and metallicity--luminosity (bottom) relations for early type galaxies, compact stellar systems, dwarf spheroidal galaxies, and ultra-diffuse galaxies. The data sources are specified in the text. The dashed line in the upper panel shows the $L \propto \sigma^4$ relation defined for massive E/S0 galaxies that continues towards compact stellar systems in the low-mass end. The Binospec measurements are shown with black dots. Black arrows indicate results of passive evolution of two younger UDGs in 10~Gyr. The red arrow points to the position of the nucleus of \emph{J130005.40+275333.0}\label{fjrmet}}
\end{figure}

\emph{J130005.40+275333.0} is a special case where triaxiality may reconcile its significant stellar rotation with its round shape. The stellar population is intermediate-age (3~Gyr) and metal-rich ($-0.3$~dex). The luminous nucleus ($m_R \approx 22.44$~mag; $M_R \approx -12.64$~mag; red arrow in Fig.~\ref{fjrmet}), is similar to the nuclei in many more luminous dE/dS0 galaxies in the Virgo cluster presented in \citet{Chilingarian09}. The nucleus ends up in the UCD loci in both Faber--Jackson and luminosity--metallicity relations.  This galaxy may be the remnant of a tidally stripped massive progenitor, with a low surface brightness envelope similar to Fornax~UCD~3 \citep{2018MNRAS.477.4856A}.

In Fig.~\ref{fjrmet} we present the Faber--Jackson (\citeyear{1976ApJ...204..668F}) and the metallicity--luminosity relations from the literature for giant \citep{1992ApJ...399..462B} and dwarf \citep{Chilingarian+08,Chilingarian09} elliptical galaxies, compact elliptical \citep{Chilingarian+09,Price+09} and ultra-compact dwarf galaxies \citep{CCB08,CMHI11}, and Local group dwarf spheroidal galaxies \citep{Mateo98}. The dataset is complemented by measurements for intermediate-luminosity and giant Coma cluster E/S0 galaxies extracted from the Reference Catalog of Spectral Energy Distributions \citep[RCSED,][]{RCSED}. All luminosities were converted into the $B$ band using transformations for quiescent galaxies from RCSED and using standard $ugriz$ to $UBVR_CI_C$ color equations.\footnote{\scriptsize \url{http://www.sdss3.org/dr8/algorithms/sdssUBVRITransform.php}} Prior to our study, there were only two velocity dispersion measurements for UDGs from integrated light spectra \citep{2016ApJ...828L...6V,Danieli+19}.  Metallicity measurements from integrated light low-resolution spectra are available for  9 Coma cluster UDGs \citep{2018ApJ...859...37G,2018MNRAS.478.2034R} and for \emph{NGC~1052-DF2} \citep{2018arXiv181207346F}. We adopt a distance of 13~Mpc for \emph{NGC~1052-DF2} \citep{2018arXiv180610141T}. 

The Faber--Jackson relation for luminous galaxies has a slope of $L \propto \sigma^4$ \citep{2003AJ....125.1849B} continuing all the way into the locus of tidally-stripped, dark-matter-free compact elliptical and ultra-compact dwarf galaxies. However, for ``normal'' dE/dS0 in clusters it changes to $L \propto \sigma^2$ at $M_B>-18.5$~mag \citep{MG05,Kourkchi+12}. Dwarf spheroidal galaxies containing substantial quantities of dark matter form a sequence almost parallel to that of compact stellar systems but a factor of 3--4 lower in velocity dispersion, before turning up at low-luminosities into the ultra-faint dwarf galaxy regime. UDGs with measured velocity dispersions fill the gap in luminosity between cluster dE/dS0s and dSph. The three most luminous objects from the \citet{Mateo98} sample in the same locus of the diagram are not dSphs but dE satellites of the Andromeda galaxy \emph{NGC~147}, \emph{NGC~185}, and \emph{NGC~205}. The black arrows pointing left indicate the direction of passive evolution of the two younger UDGs in our sample to the age of 10~Gyr computed using {\sc pegase.hr} stellar population models. Passive evolution will bring \emph{J130026.26+272735.2} to the middle of the UDG locus.

In the luminosity--metallicity relation UDGs also fall on the monotonic relation formed by dSph, dE/dS0, and E/S0 galaxies extending over a range of 150,000 in $B$-band luminosity (13~mag). The strongest outlier is \emph{J130005.40+275333.0} where the unresolved stellar nucleus is even more metal rich, making it similar to some cluster dE/dS0 galaxies \citep{Chilingarian09}. The luminosity--metallicity relation is commonly thought to result from supernovae driven winds \citep{1986ApJ...303...39D}. Under this assumption, UDGs are well placed in the landscape formed by early-type galaxies of larger (dE/dS0) and smaller (dSph) stellar masses.

The 9 observed UDGs form a heterogeneous population similar to brighter dwarf elliptical galaxies in clusters identified by \citet{2007ApJ...660.1186L}. Some UDGs rotate, some do not, and one has a stellar nucleus even though the fraction of nucleated dwarf elliptical galaxies drops at low luminosities. We see at least one object with a peculiar morphology, where we suspect tidal interaction with a more massive dwarf elliptical companion. All 9 UDGs appear to be less luminous and lower surface brightness analogs of existing sub-classes of dE/dS0s, approaching the properties of Local Group dSph. Our observations suggest that UDGs are a fainter extension of the dE/dS0 class rather than a distinct galaxy type, as proposed by \citet{2018RNAAS...2a..43C}.

A diversity of formation scenarios can explain dwarf elliptical galaxies in clusters and groups, e.g. supernovae feedback \citep{1986ApJ...303...39D}, ram pressure stripping \citep{1972ApJ...176....1G}, numerous tidal encounters with more massive cluster members during flybys \citep{1996Natur.379..613M}. Most of these scenarios explain the morphological transformation of gas rich disky progenitors into quiescent regularly shaped dE/dS0 galaxies. These processes cause disk heating and might also boost the orbital anisotropy if a substantial mass fraction in a form of gas is rapidly stripped. Even in our very limited sample of 9 UDGs, we can see examples of galaxies that possibly formed in each of the three scenarios. However, dwarf triaxial systems such as \emph{J130005.40+275333.0} and \emph{VCC~1545} cannot be explained by these scenarios and present a challenge to theories of galaxy formation, because their giant counterparts are usually explained as results of rich merger histories \citep{2001ApJ...548L..33D} inconsistent with low-mass systems.

\acknowledgments

We thank the Binospec commissioning team at SAO and at the MMT. IC and SM are supported by Telescope Data Center at SAO. IC, KG, and AA acknowledge support from the Russian Science Foundation grant 17-72-20119 and the Leading Scientific School in astrophysics (direction: extragalactic astronomy) at M.~V.~Lomonosov Moscow State University. KG also acknowledges support by the Basis foundation stipend (category: students).

%

\facilities{MMT, Subaru}





\bibliographystyle{aasjournal}
\bibliography{UDG}



\end{document}